\documentclass{aastex631}
\usepackage{enumerate}

\usepackage{amssymb, amsmath}

\usepackage{natbib}
\usepackage{xcolor}
\usepackage{ulem}
\usepackage{dblfnote}
\usepackage{appendix}
\usepackage[stable]{footmisc}
\usepackage{comment}
\usepackage{float}
\usepackage{array}
\usepackage{booktabs}
\usepackage[a4paper, top=1.5in, bottom=1in, left=1.25in, right=.75in]{geometry}

\usepackage[version=4]{mhchem}
\usepackage{graphicx}

\usepackage{hyperref}
\begin{document}

\title{Enhancing Exoplanet Ephemerides by Leveraging Professional and Citizen Science Data: A Test Case with WASP-77A b}

\author[0000-0003-4295-7313]{Federico R. Noguer}[*]
\affiliation{School of Earth and Space Exploration, Arizona State University, 781 E Terrace Mall, Tempe, AZ 85287-6004, United States}

\author{Suber Corley}
\affiliation{School of Earth and Space Exploration, Arizona State University, 781 E Terrace Mall, Tempe, AZ 85287-6004, United States}

\author[0000-0002-5785-9073]{Kyle A. Pearson}
\affiliation{Jet Propulsion Laboratory California Institute of Technology. 4800 Oak Grove Drive, Pasadena, CA 91109, USA}

\author[0000-0001-7547-0398]{Robert T. Zellem}
\affiliation{Jet Propulsion Laboratory California Institute of Technology. 4800 Oak Grove Drive, Pasadena, CA 91109, USA}

\author[0000-0001-8536-0942]{Molly N. Simon}
\affiliation{School of Earth and Space Exploration, Arizona State University, 781 E Terrace Mall, Tempe, AZ 85287-6004, United States}
%\collaboration{20}{(Eager Beaver Research)}

\author[0000-0002-0040-6815]{Jennifer A. Burt}
\affiliation{Jet Propulsion Laboratory California Institute of Technology. 4800 Oak Grove Drive, Pasadena, CA 91109, USA}

\author{Isabela Huckabee}
\affiliation{School of Earth and Space Exploration, Arizona State University, 781 E Terrace Mall, Tempe, AZ 85287-6004, United States}
\affiliation{Jet Propulsion Laboratory California Institute of Technology. 4800 Oak Grove Drive, Pasadena, CA 91109, USA}

\author[0000-0003-3829-8554]{Prune C. August}
\affiliation{DTU Space, Technical University of Denmark, Elektrovej 328, DK-2800 Kgs. Lyngby, Denmark}

\author[0000-0003-4241-7413]{Megan Weiner Mansfield}
\affiliation{Steward Observatory, University of Arizona, Tucson, AZ 85719, USA}
\affiliation{NHFP Sagan Fellow}

\author[0000-0002-4297-5506]{Paul A. Dalba} 
\affiliation{Department of Astronomy and Astrophysics, University of California, Santa Cruz, CA 95064, USA}
\affiliation{SETI Institute, Carl Sagan Center, 339 Bernardo Ave, Suite 200, Mountain View, CA 94043, USA}

\author[0000-0002-9946-5259]{Peter C. B. Smith}
\affiliation{School of Earth $\&$ Space Exploration, Arizona State University, Tempe AZ 85287, USA}

\author[0000-0001-9445-4588]{Timothy Banks}
\affiliation{Exoplanet Watch Citizen Scientist Contributor, https://exoplanets.nasa.gov/exoplanet-watch}
\affiliation{Dept. of Physical Sci. \& Engineering, Harper College, 1200 W Algonquin Rd, Palatine, IL 60067, USA}
\affiliation{Nielsen, 675 6th Ave, New York 10011, USA}

\author{Ira Bell}
\affiliation{School of Earth and Space Exploration, Arizona State University, 781 E Terrace Mall, Tempe, AZ 85287-6004, United States}
\affiliation{Exoplanet Watch Citizen Scientist Contributor, https://exoplanets.nasa.gov/exoplanet-watch}

\author{Dominique  Daniel}
\affiliation{ExoClock Project Citizen Scientist Contributor, https://www.exoclock.space}

\author{Lindsay Dawson}
\affiliation{Thomas More University, 333 Thomas More Parkway, Crestview Hills, Kentucky}
\affiliation{Exoplanet Watch Citizen Scientist Contributor, https://exoplanets.nasa.gov/exoplanet-watch}

\author{Jesús De Mula}
\affiliation{Exoplanet Watch Citizen Scientist Contributor, https://exoplanets.nasa.gov/exoplanet-watch}

\author{Marc Deldem}
\affiliation{ExoClock Project Citizen Scientist Contributor, https://www.exoclock.space}

\author{Dimitrios Deligeorgopoulos}
\affiliation{ExoClock Project Citizen Scientist Contributor, https://www.exoclock.space}

\author{Romina P. Di Sisto}
\affiliation{Facultad de Ciencias Astronómicas y Geofísicas Universidad Nacional de La Plata, Argentina}
\affiliation{Instituto de Astrofísica de La Plata, Observatorio Astronómico de La Plata: Paseo del Bosque, B1900FWA, La Plata, Argentina}

\author{Roger Dymock }
\affiliation{British Astronomical Association}

\author[0000-0002-5674-2404]{Phil Evans}
\affiliation{El Sauce Observatory, Coquimbo Province, Chile}

\author{Giulio Follero}
\affiliation{Unione Astrofili Napoletani, c/o INAF-Osservatorio Astronomico di Capodimonte Napoli via Moiariello, 16 80131 Napoli}

\author[0000-0003-4211-3671]{Martin J. F. Fowler}
\affiliation{Les Rocquettes Observatory and Exoplanet Factory, South Wonston, Hampshire, SO21 3EX, United Kingdom}

\author{Eduardo Fernández-Lajús }
\affiliation{Instituto de Astrofísica de La Plata, Observatorio Astronómico de La Plata: Paseo del Bosque, B1900FWA, La Plata, Argentina}

\author{Alex Hamrick}
\affiliation{Stanford Online High School, 415 Broadway Academy Hall, Floor 2, 8853, Redwood City, CA 94063}

\author{Nicoletta Iannascoli}
\affiliation{Unione Astrofili Napoletani, c/o INAF-Osservatorio Astronomico di Capodimonte Napoli via Moiariello, 16 80131 Napoli}

\author{Andre O. Kovacs}
\affiliation{AAVSO, 185 Alewife Brook Parkway, Suite 410, Cambridge, MA 02138, USA}
\affiliation{Exoplanet Watch Citizen Scientist Contributor, https://exoplanets.nasa.gov/exoplanet-watch}

\author{Denis Henrique Kulh}
\affiliation{Alfa Crucis}

\author{Claudio Lopresti}
\affiliation{ExoClock Project Citizen Scientist Contributor, https://www.exoclock.space}
\affiliation{Gruppo Astronomia Digitale (GAD, Italy)}
\affiliation{Sezione Pianeti Extrasolari (UAI, Italy)}

\author{Antonio Marino}
\affiliation{Unione Astrofili Napoletani, c/o INAF-Osservatorio Astronomico di Capodimonte Napoli via Moiariello, 16 80131 Napoli}

\author{Bryan E. Martin}
\affiliation{Tiger Butte Observatory, 2013 Whispering Ridge Dr, Great Falls, MT 59405.}
\affiliation{AAVSO, 185 Alewife Brook Parkway, Suite 410, Cambridge, MA 02138, USA}

\author{Paolo Arcangelo Matassa}
\affiliation{ExoClock Project Citizen Scientist Contributor, https://www.exoclock.space}

\author{Tasso Augusto Napoleão}
\affiliation{Alfa Crucis}

\author{Alessandro Nastasi}
\affiliation{GAL Hassin - Centro Internazionale per le Scienze Astronomiche, Via della Fontana Mitri 3, 90010 Isnello, Palermo, Italy}
\affiliation{INAF - Osservatorio Astronomico di Palermo, Piazza del Parlamento, 1, 90134 Palermo, Italy}

\author{Anthony Norris}
\affiliation{Exoplanet Watch Citizen Scientist Contributor, https://exoplanets.nasa.gov/exoplanet-watch}

\author{Alessandro Odasso }
\affiliation{ExoClock Project Citizen Scientist Contributor, https://www.exoclock.space}

\author{ Nikolaos I. Paschalis}
\affiliation{Nunki Observatory, Skiathos Island, Greece}
\affiliation{ExoClock Project Citizen Scientist Contributor, https://www.exoclock.space}

\author{ Pavel  Pintr}
\affiliation{Institute of Plasma Physics ASCR, v. v. i. Centre TOPTEC, Sobotecka 1660 511 01 Turnov Czech Republic}
\affiliation{ExoClock Project Citizen Scientist Contributor, https://www.exoclock.space}

\author{Jake Postiglione}
\affiliation{Exoplanet Watch Citizen Scientist Contributor, https://exoplanets.nasa.gov/exoplanet-watch}

\author{Justus Randolph}
\affiliation{Exoplanet Watch Citizen Scientist Contributor, https://exoplanets.nasa.gov/exoplanet-watch}

\author{François Regembal}
\affiliation{ExoClock Project Citizen Scientist Contributor, https://www.exoclock.space}

\author{Lionel Rousselot}
\affiliation{ExoClock Project Citizen Scientist Contributor, https://www.exoclock.space}

\author{Sergio José Gonçalves da Silva}
\affiliation{Alfa Crucis}

\author{Andrew Smith}
\affiliation{Exoplanet Watch Citizen Scientist Contributor, https://exoplanets.nasa.gov/exoplanet-watch}

\author{Andrea Tomacelli}
\affiliation{Unione Astrofili Napoletani, c/o INAF-Osservatorio Astronomico di Capodimonte Napoli via Moiariello, 16 80131 Napoli}

\section{Abstract} \label{abstract}
We present an updated ephemeris, and physical parameters, for the exoplanet WASP-77 A b. In this effort, we combine 64 ground- and space-based transit observations, 6 space-based eclipse observations, and 32 radial velocity observations to produce this target’s most precise orbital solution to date aiding in the planning of 
James Webb Space Telescope (JWST) and Ariel observations and atmospheric studies.  We report a new orbital period of $1.360029395 \pm 5.7 \times 10 ^{-8}$ days, a new mid-transit time of $2459957.337860 \pm 4.3 \times 10 ^{-5}$ BJDTDB (Barycentric Julian Date in the Barycentric Dynamical Time scale; \citet{TDB_Eastman_2010PASP..122..935E}) and a new mid-eclipse time of $2459956.658192  \pm6.7\times10^{-5}$ BJDTDB. Furthermore, the methods presented in this study reduce the uncertainties in the planet’s mass 1.6654 $\pm4.5\times10^{-3} M_{Jup}$ and orbital period $1.360029395 \pm 5.7 \times 10 ^{-8}$ days by factors of 15.1 and 10.9, respectively. Through a joint fit analysis comparison of transit data taken by space-based and citizen science-led initiatives, our study demonstrates the power of including data collected by citizen scientists compared to a fit of the space-based data alone.  Additionally, by including a vast array of citizen science data from ExoClock, Exoplanet Transit Database (ETD), and Exoplanet Watch, we can increase our observational baseline and thus acquire better constraints on the forward propagation of our ephemeris than what is achievable with TESS data alone.


\section{Introduction} \label{sec:intro}

Correctly determining a planet’s transit ephemeris is essential for the efficient scheduling of follow-up observations \citep{zellem2020} with telescopes such as the James Webb Space Telescope (JWST), the Hubble Space Telescope (HST), or the upcoming Ariel mission scheduled to launch in 2029 that will observe the atmospheres of 1000 exoplanets. Accurate ephemerides help ensure that these highly competitive space-based telescopes are utilized as efficiently as possible to maximize their science output by helping to reduce overheads . Precise exoplanet mid-transit times, whereby a planet passes directly in front of its host star, are crucial for characterizing an exoplanet's atmosphere through transmission spectroscopy. Due to ephemeris uncertainties, observation time buffers must be included before and after the predicted mid-transit time to ensure that the entire transit is captured in addition to pre and post transit baseline measurements of just the host star flux. As a planet completes additional orbits around its host star, uncertainties in its mid-transit time increase, creating the need for repeated follow-up observations and new analyses \citep{dragomir2019,zellem2020,kokori2023}. Since the number of confirmed exoplanets and exoplanet candidates continues to increase, expected to surpass 10,000 from NASA’s TESS mission alone \citep{Kunimoto2022}, there will be a high demand for updated exoplanet ephemerides to enable more efficient analyses of these new worlds.

Previous studies have demonstrated that successfully updating an exoplanet’s mid-transit time can be achieved through collaborations with professionals and citizen scientists utilizing small, ground-based telescopes  \citep[see, e.g.,][]{mizrachi202125,fowler2021JBAA..131..359F,pearson2022,hewitt2023}. In one such example, students enrolled in a fully online research course at Arizona State University acquired observations of the planet WASP-104 b taken with a ground-based, robotic, 6-inch telescope \citep{hewitt2023,hewitt2023course}. These students were able to improve upon the uncertainty on the mid-transit time for this particular target by 2.7 percent when compared to the next most recent publication. Efforts like these lay the foundation for continued professional and citizen science-based collaborations that can assist with exoplanet ephemeris maintenance. 

Along with a planet’s mid-transit time, the mid-eclipse time is also a crucial parameter for astronomers characterizing exoplanetary atmospheres. During eclipse, the planet passes behind the star, temporarily blocking its reflected light and thermal emission. Eclipses, when the planet passes behind its host star, enable astronomers to measure a planet's reflected light and thermal emission. For example \cite{smith2024combined} leveraged this method with JWST to use the difference in the observed light curve of WASP-77 A b during eclipse, as compared to just before or after, to isolate the planet’s own emission from that of the star. Determining a planet's mid-eclipse time when an eclipse was not previously observed, requires not only its mid-transit time, but also knowledge of the planet’s eccentricity and argument of periastron, parameters that can be deduced from radial velocity measurements. As such, the inclusion of radial velocity data to the analysis significantly improves precision in predicting transit, eclipse timing \citep{pearson2022}, and the planet's orbit for the planning of JWST and Ariel observations \citep{zellem2020,burtzellem2024}.  Additionally, constraints on the planet's mass helps alleviate degeneracies when performing atmospheric fits \citep{batalha2018}.

In this effort, we use WASP-77 A b to demonstrate the power of combining a broad suite of transit and eclipse photometry data (collected by both citizen scientists and space-based telescopes) with archival radial velocity data to acquire precise orbital and planetary parameters. WASP-77 A b, a hot Jupiter orbiting a G8 V spectral-type star with V-mag = 10.294 ± 0.007 \citep{maxted2013}.  

WASP-77 A b has been the target of multiple spectroscopic campaigns designed to measure its atmospheric composition and constrain its formation history. Via high resolution spectroscopic measurements with Gemini-South/IGRINS, WASP-77 A b's atmosphere was initially inferred to have a subsolar metallicity and solar C/O ratio \citep{line2021solar}, an uncommon but not implausible prediction within the core accretion paradigm \citep[e.g.,][]{madhusudhan2017atmospheric}. Measurements with HST's Wide Field Camera 3 (WFC3) favored instead a supersolar metallicity \citep{mansfield2022}, in apparent tension with the previous subsolar inference from Gemini/IGRINS. However, recent observations with JWST/NIRSpec \citep{august2023} as well as a joint analysis of both the NIRSpec and IGRINS data \citep{smith2024combined}, have confirmed the original subsolar metallicity measurement. Additionally, a recent reanalysis of the WFC3 data that takes into account contamination of the stellar companion WASP-77 B found that the WFC3 data indeed also favors a subsolar metallicity \citep{edwards2024measuring}. Within the context of the formation simulations by \cite{bitsch2022drifting,khorshid2023retrieving}, WASP-77 A b's atmospheric composition is consistent with formation beyond the CO$_{2}$ ice line and inward migration after disk dissipation. The planet star system does not reveal evidence of rapid orbital decay nor other significant orbital perturbations \citep{cortes2020tramos}. Currently, no additional planetary companions have been reported for WASP 77 A and upper mass limits for potential companions have been calculated by \cite{cortes2020tramos}. 

WASP-77 A b is an ideal candidate for this study due the significant amount of archival transit and radial velocity data readily available for this target. As such, to perform the combined transit and radial velocity analysis described in the remainder of this paper, we collected 34 transit photometry data from three different citizen science initiatives: ExoClock \citep{kokori2022exoclockdata1,kokori2022exoclockdata2,kokori2023exoclockdata3}, Exoplanet Watch \citep{zellem2020}, and the Exoplanet Transit Database (ETD)\citep{ETD2010} along with space-based photometry data from two sectors of TESS \citep{ricker2015tess}. These mid-transit measurements were combined with archival radial velocity data from the CORALIE and High Accuracy Radial velocity Planet Searcher (HARPS) spectrographs in conjunction with mid-eclipse times derived from JWST, Spitzer and HST. The analysis and methods implemented in this work for WASP-77 A b can be applied when updating the ephemerides of other exoplanets that are high priority targets \citep[e.g.,][]{zellem2017forecasting,kempton2018framework, hord2023identification} for follow-up analysis with larger ground-based or space-based initiatives.

\section{OBSERVATIONS} \label{sec:obs}

\subsection{Citizen Science Data} 
\label{subsec:cit}
\begin{figure}[H]
\includegraphics[width=0.9\textwidth]{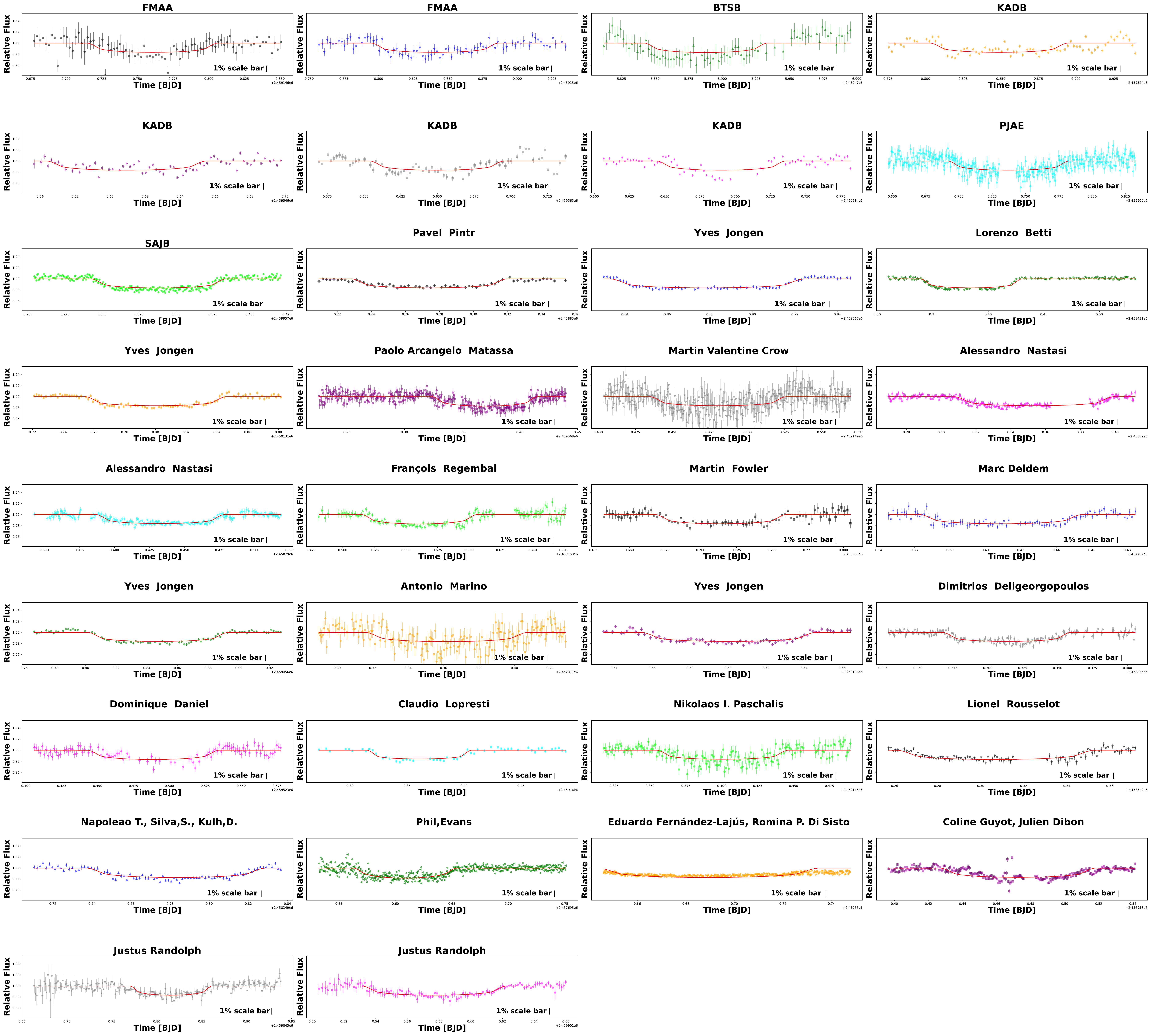}
\caption{Individual light curves were produced from 34 sets of terrestrial citizen science data during the process described in \nameref{sec:transitreduction} Section \ref{sec:transitreduction}. Citizen science observations span approximately 8 years from October, 2014 to January, 2023.}  
\label{fig:citiindividualplots}
\end{figure}

We incorporated 34 observations from citizen science initiatives (11 from Exoplanet Watch, Table \ref{tab:Exoplanet Watch Single Fit Outputs}; 19 from ExoClock, Table \ref{tab:exoclocktmids}; 4 from the Exoplanet Transit Database, Table \ref{tab:etdtmids}) taken with telescopes as small as 4.5-inch Unistellar \citep{peluso2023unistellar} eVscopes and the Harvard Smithsonian Center for Astrophysics' 6-inch ``Cecilia'' MicroObservatory Robotic Telescope on Mount Hopkins in Arizona \citep{sadler2001microobservatory}.
Citizen science networks can help refresh and improve transit ephemerides at far less cost than the equivalent effort of larger and more competitive professionally operated facilities \citep{zellem2020}.  For example, the Exoclock project \citep{kokori2021,kokori2022exoclockdata1,kokori2022exoclockdata2,kokori2023exoclockdata3} has updated the ephemerides for at least 450 planets using space-based, large ground-based and citizen-science observations from the Exoclock network. 
The data obtained by citizen scientists is made publicly available to the scientific community.  Exoplanet Watch citizen science observations are uploaded to AAVSO and can also be accessed through the Exoplanet Watch website or programmatically with EXOTIC code. Exoclock and ETD also have depositories for their citizen transit data that can be publicly accessed. 
Figure \ref{fig:citiindividualplots} displays the resulting light curves from all 34 citizen science transit observations.

\begin{deluxetable}{lllll}[H]
\tabletypesize{\scriptsize}
\centering
\tablecaption{WASP-77~A~b Derived Parameters from Exoplanet Watch Transit Data Individual Fits \label{tab:Exoplanet Watch Single Fit Outputs}}
\tablehead{
\colhead{Observer} & \colhead{Rp/Rs} & \colhead{Mid-transit (BJDTDB)} & \colhead{Inclination (deg.)} & \colhead{$\boldsymbol{\sigma_{res}}$}}
\startdata
\hline
Martin J. F. Fowler (FMAA) (this study)& 0.138488 $\pm$ 0.011202 & 2459146.757993 $\pm$ 0.003096 & 85.313903 $\pm$ 1.445707 & 0.011491 \\
Martin J. F. Fowler (FMAA) (this study) & 0.130407 $\pm$ 0.008210 & 2459150.837953 $\pm$ 0.001785 & 84.937192 $\pm$ 1.059638 & 0.006623 \\
BTSB (this study)& 0.167188 $\pm$ 0.009166 & 2459475.883336 $\pm$ 0.002862 & 89.775798 $\pm$ 1.057500 & 0.010702 \\
Andre O. Kovacs (KADB) (this study)& 0.109212 $\pm$ 0.002484 & 2459524.851427 $\pm$ 0.000618 & 89.970091 $\pm$ 0.490967 & 0.008491 \\
Andre O. Kovacs (KADB) (this study)& 0.109670 $\pm$ 0.003064 & 2459546.613122 $\pm$ 0.000767 & 89.863576 $\pm$ 0.606149 & 0.007160 \\
Andre O. Kovacs (KADB) (this study)& 0.133428 $\pm$ 0.004877 & 2459565.643618 $\pm$ 0.001494 & 87.404846 $\pm$ 1.337892 & 0.009443 \\
Andre O. Kovacs (KADB) (this study)& 0.145692 $\pm$ 0.001675 & 2459584.691209 $\pm$ 0.000363 & 84.910237 $\pm$ 0.022267 & 0.006784 \\
Jake Postiglione (PJAE) (this study)& 0.140661 $\pm$ 0.005349 & 2459909.739026 $\pm$ 0.001112 & 87.120092 $\pm$ 1.155684 & 0.009801 \\
SAJB (this study)& 0.133104 $\pm$ 0.000757 & 2459957.337524 $\pm$ 0.000198 & 88.615123 $\pm$ 0.497671 & 0.003242 \\
Justus Randolph (this study)& 0.114845 $\pm$ 0.004284 & 2459845.810377 $\pm$ 0.001304 & 89.984181 $\pm$ 0.762004 & 0.006574 \\
Justus Randolph (this study)& 0.121547 $\pm$ 0.003251 & 2459901.578087 $\pm$ 0.001122 & 89.053012 $\pm$ 1.047644 & 0.004139 \\
\enddata
\tablecomments{The observations in this table were acquired from NASA Exoplanet Watch citizen science project \citep{zellem2020}}
\end{deluxetable}

\begin{deluxetable}{lllll}[H]
\tabletypesize{\scriptsize}
\centering
\tablecaption{WASP-77~A~b Derived Parameters from ExoClock Transit Data Individual Fits\label{tab:exoclocktmids}}
\tablehead{
\colhead{Observer} & \colhead{Rp/Rs} & \colhead{Mid-transit (BJD)} & \colhead{Inclination (deg.)} & \colhead{$\boldsymbol{\sigma_{res}}$}}
\startdata
\hline
 Pavel  Pintr (this study)
 & 0.100245 $\pm$ 0.003589 & 2458850.274405 $\pm$ 0.000707 & 88.273883 $\pm$ 0.907354 & 0.001845 \\
 Yves  Jongen (this study)
 & 0.129250 $\pm$ 0.001734 & 2459067.879321 $\pm$ 0.000514 & 89.150409 $\pm$ 0.641214 & 0.001676 \\
 Lorenzo  Betti (this study)
 & 0.129802 $\pm$ 0.001138 & 2458431.383446 $\pm$ 0.000269 & 87.606257 $\pm$ 0.405009 & 0.001453 \\
 Yves  Jongen (this study)
 & 0.124696 $\pm$ 0.002344 & 2459131.799264 $\pm$ 0.000654 & 89.448679 $\pm$ 0.781300 & 0.002639 \\
 Paolo Arcangelo  Matassa (this study)
 & 0.132062 $\pm$ 0.003724 & 2459568.375690 $\pm$ 0.002321 & 84.917774 $\pm$ 1.421284 & 0.007001 \\
 Martin Valentine Crow (this study)
 & 0.140826 $\pm$ 0.007180 & 2459149.479783 $\pm$ 0.001457 & 85.368562 $\pm$ 1.203334 & 0.014618 \\
 Alessandro  Nastasi (this study)
 & 0.115978 $\pm$ 0.002669 & 2458820.353259 $\pm$ 0.000452 & 89.886516 $\pm$ 0.457542 & 0.003295 \\
 Alessandro  Nastasi (this study)
 & 0.112150 $\pm$ 0.003711 & 2458790.433805 $\pm$ 0.000826 & 87.571683 $\pm$ 1.030419 & 0.003678 \\
 François  Regembal (this study)
 & 0.128294 $\pm$ 0.003477 & 2459153.562620 $\pm$ 0.000698 & 87.453267 $\pm$ 0.964544 & 0.005479 \\
 Martin  Fowler (this study)
 & 0.116704 $\pm$ 0.006113 & 2458855.715261 $\pm$ 0.001465 & 89.997814 $\pm$ 1.132437 & 0.005860 \\
 Marc Deldem (this study)
 & 0.122177 $\pm$ 0.004267 & 2457702.409628 $\pm$ 0.001268 & 89.991455 $\pm$ 0.845319 & 0.005132 \\
 Yves  Jongen (this study)
 & 0.131871 $\pm$ 0.002423 & 2459456.847265 $\pm$ 0.000472 & 87.687640 $\pm$ 0.792568 & 0.002031 \\
 Antonio  Marino (this study)
 & 0.150334 $\pm$ 0.007487 & 2457377.366686 $\pm$ 0.001927 & 87.237357 $\pm$ 1.361995 & 0.014233 \\
 Yves  Jongen (this study)
 & 0.130505 $\pm$ 0.003079 & 2459138.600169 $\pm$ 0.000722 & 89.893542 $\pm$ 0.743697 & 0.003175 \\
 Dimitrios  Deligeorgopoulos (this study)
 & 0.115777 $\pm$ 0.003506 & 2458835.313724 $\pm$ 0.000657 & 87.954140 $\pm$ 0.924682 & 0.004149 \\
 Dominique  Daniel (this study)
 & 0.127054 $\pm$ 0.007057 & 2459523.492783 $\pm$ 0.001510 & 84.981891 $\pm$ 1.009943 & 0.006525 \\
 Claudio  Lopresti (this study)
 & 0.129025 $\pm$ 0.002699 & 2459160.361783 $\pm$ 0.000740 & 86.890974 $\pm$ 0.908180 & 0.002571 \\
 Nikolaos I. Paschalis (this study)
 & 0.135951 $\pm$ 0.006095 & 2459145.399718 $\pm$ 0.001569 & 89.758226 $\pm$ 1.279541 & 0.009116 \\
 Lionel  Rousselot (this study)
 & 0.119026 $\pm$ 0.004222 & 2458529.309950 $\pm$ 0.000921 & 89.810029 $\pm$ 0.746735 & 0.003881 \\
\enddata
\tablecomments{The observations in this table were acquired from the ExoClock citizen science project \citep{kokori2022exoclockdata1,kokori2022exoclockdata2,kokori2023exoclockdata3}}
\end{deluxetable}

\begin{deluxetable}{lllll}[H]
\tabletypesize{\scriptsize}
\centering
\tablecaption{WASP-77~A~b Derived Parameters from the Exoplanet Transit Database (ETD) Transit Data Individual Fits\label{tab:etdtmids}}
\tablehead{
\colhead{Observer} & \colhead{Rp/Rs} & \colhead{Mid-transit (BJDTDB)} & \colhead{Inclination (deg.)} & \colhead{$\boldsymbol{\sigma_{res}}$}}
\startdata
\hline
 Napoleao T., Silva,S., Kulh,D.
 (this study)& 0.125011 $\pm$ 0.002695 & 2458349.782454 $\pm$ 0.000589 & 87.631431 $\pm$ 0.892017 & 0.003467 \\
 Phil,Evans
 (this study)& 0.112988 $\pm$ 0.000419 & 2457695.608264 $\pm$ 0.000147 & 89.978600 $\pm$ 0.109655 & 0.005567 \\
 Fernández-Lajús E., Di Sisto R. P.
 (this study)& 0.069934 $\pm$ 0.001692 & 2459550.688000 $\pm$ 0.000443 & 85.250493 $\pm$ 0.222586 & 0.001769 \\
 Coline Guyot, Julien Dibon
 (this study)& 0.119733 $\pm$ 0.001505 & 2456958.477899 $\pm$ 0.000568 & 84.910753 $\pm$ 0.043606 & 0.005393 \\
\enddata
\tablecomments{Observations in this table were acquired from the Exoplanet Transit Database (ETD) \citep{ETD2010}}
\end{deluxetable}

\subsection{Transiting Exoplanet Survey Satellite (TESS) Data} \label{subsec:tess}
\begin{figure}[H]
\includegraphics[width=0.9\textwidth]{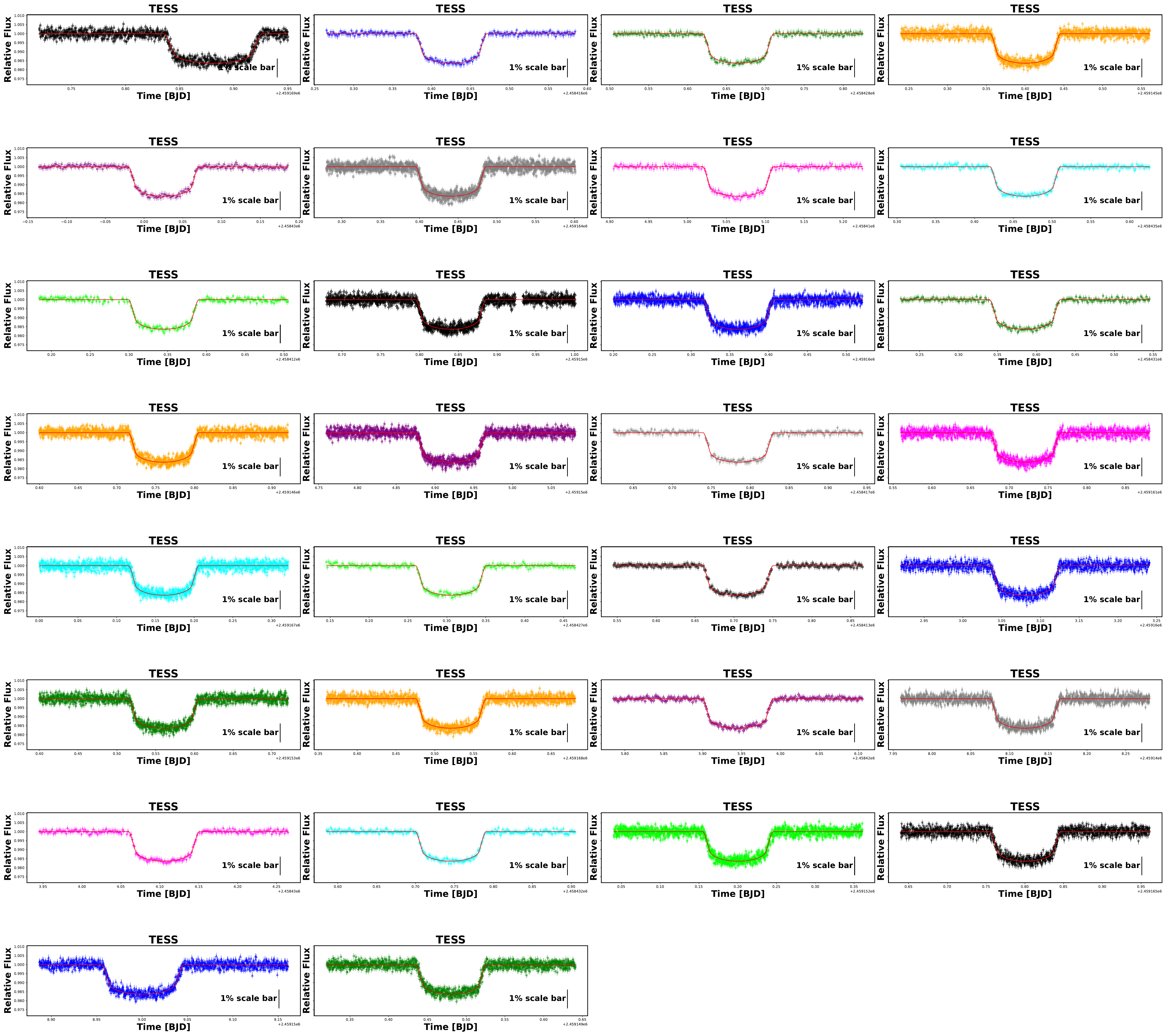}
\caption{These individual light curves are produced from the 30 TESS transit observations during the process described in \nameref{sec:transitreduction} Section \ref{sec:transitreduction}. The wellness of fit in each light curve demonstrates the reliability of TESS for precise transit photometry measurements.TESS observations span approximately 2 years from October, 2018 to November, 2020.}
\label{fig:tessindividualplots}
\end{figure}

WASP-77 A was observed by TESS during Sector 4 (2018-10-19 to 2018-11-14) and Sector 31 (2020-10-22 to 2020-11-18) in 2 minute cadence during which time WASP-77 A b completed 30 transits (Table \ref{tab:TESS Fits}). The individual time series files used in this study were obtained from public TESS data on the Mikulski Archive for Space Telescopes (MAST) following the prescription described in \cite{pearson2019}, which performs an updated aperture selection and can improve the photometric scatter by up to 30\% for some targets by reducing pointing-induced systematics. We removed any data points taken within 30 minutes of the TESS momentum dumps, as these data points typically contain large systematic errors. We next removed 3 sigma outliers from a rolling median filter with a bin size of 32 minutes, as well as any quality-flagged data points. We also applied the Python package \texttt{WOTAN} \citep{HippkeWOTAN} to detrend WASP-77 A's variability from the light curves while ensuring the transit signal remains intact. Figure~\ref{fig:tessindividualplots} displays the resulting light curves.

\begin{deluxetable}{lllll}[H]
\tabletypesize{\scriptsize}
\centering
\tablecaption{WASP-77~A~b Derived Parameters from the Transiting Exoplanet Survey Satellite (TESS) Transit Data Individual Fits\label{tab:TESS Fits}}
\tablehead{
\colhead{Source} & \colhead{Rp/Rs} & \colhead{Mid-transit (BJDTDB)} & \colhead{Inclination (deg.)} & \colhead{$\boldsymbol{\sigma_{res}}$}}
\startdata
TESS (this study) & 0.117211 $\pm$ 0.000484 & 2459169.880555 $\pm$ 0.000141 & 88.173852 $\pm$ 0.290894 & 0.001788 \\
TESS (this study)& 0.118047 $\pm$ 0.001131 & 2458416.424934 $\pm$ 0.000362 & 87.712535 $\pm$ 0.633352 & 0.000714 \\
TESS (this study)& 0.117444 $\pm$ 0.001152 & 2458428.664764 $\pm$ 0.000363 & 87.963218 $\pm$ 0.692152 & 0.000741 \\
TESS (this study)& 0.117935 $\pm$ 0.000484 & 2459145.400138 $\pm$ 0.000142 & 88.168883 $\pm$ 0.277498 & 0.001665 \\
TESS (this study)& 0.117839 $\pm$ 0.001126 & 2458430.025059 $\pm$ 0.000353 & 87.884216 $\pm$ 0.671382 & 0.000796 \\
TESS (this study)& 0.117399 $\pm$ 0.000466 & 2459164.440698 $\pm$ 0.000142 & 88.106095 $\pm$ 0.259252 & 0.001814 \\
TESS (this study)& 0.118875 $\pm$ 0.001119 & 2458415.064802 $\pm$ 0.000346 & 87.598829 $\pm$ 0.530363 & 0.000812 \\
TESS (this study)& 0.116758 $\pm$ 0.001099 & 2458435.465274 $\pm$ 0.000339 & 88.042295 $\pm$ 0.664903 & 0.000765 \\
TESS (this study)& 0.117346 $\pm$ 0.001158 & 2458412.344649 $\pm$ 0.000349 & 88.291218 $\pm$ 0.677087 & 0.000905 \\
TESS (this study)& 0.117514 $\pm$ 0.000461 & 2459150.840086 $\pm$ 0.000140 & 87.934221 $\pm$ 0.220790 & 0.001703 \\
TESS (this study)& 0.117219 $\pm$ 0.000466 & 2459160.360695 $\pm$ 0.000145 & 87.852512 $\pm$ 0.218656 & 0.001622 \\
TESS (this study)& 0.118470 $\pm$ 0.001079 & 2458431.384909 $\pm$ 0.000333 & 87.716047 $\pm$ 0.584771 & 0.000796 \\
TESS (this study)& 0.117228 $\pm$ 0.000466 & 2459146.760236 $\pm$ 0.000141 & 88.069922 $\pm$ 0.253357 & 0.001665 \\
TESS (this study)& 0.118145 $\pm$ 0.000446 & 2459154.920533 $\pm$ 0.000139 & 87.800839 $\pm$ 0.207198 & 0.001666 \\
TESS (this study)& 0.117360 $\pm$ 0.001118 & 2458417.784887 $\pm$ 0.000360 & 87.782367 $\pm$ 0.646523 & 0.000781 \\
TESS (this study)& 0.118147 $\pm$ 0.000438 & 2459161.720527 $\pm$ 0.000139 & 87.843743 $\pm$ 0.211657 & 0.001676 \\
TESS (this study)& 0.115917 $\pm$ 0.000477 & 2459167.160783 $\pm$ 0.000151 & 88.613443 $\pm$ 0.409135 & 0.001725 \\
TESS (this study)& 0.116917 $\pm$ 0.001061 & 2458427.305095 $\pm$ 0.000358 & 88.297732 $\pm$ 0.686959 & 0.000794 \\
TESS (this study)& 0.117727 $\pm$ 0.001095 & 2458413.705108 $\pm$ 0.000354 & 87.799733 $\pm$ 0.634761 & 0.000708 \\
TESS (this study)& 0.117967 $\pm$ 0.000487 & 2459163.080668 $\pm$ 0.000148 & 88.139851 $\pm$ 0.278206 & 0.001715 \\
TESS (this study)& 0.118133 $\pm$ 0.000464 & 2459153.560254 $\pm$ 0.000142 & 88.156333 $\pm$ 0.274597 & 0.001642 \\
TESS (this study)& 0.118279 $\pm$ 0.000463 & 2459168.520838 $\pm$ 0.000139 & 87.931649 $\pm$ 0.222379 & 0.001666 \\
TESS (this study)& 0.116907 $\pm$ 0.001010 & 2458425.945066 $\pm$ 0.000330 & 88.750602 $\pm$ 0.613404 & 0.000774 \\
TESS (this study)& 0.116926 $\pm$ 0.000475 & 2459148.120307 $\pm$ 0.000140 & 87.840918 $\pm$ 0.200418 & 0.001633 \\
TESS (this study)& 0.118292 $\pm$ 0.001041 & 2458434.105215 $\pm$ 0.000335 & 88.173898 $\pm$ 0.681437 & 0.000770 \\
TESS (this study)& 0.117624 $\pm$ 0.001107 & 2458432.745412 $\pm$ 0.000348 & 88.002729 $\pm$ 0.679718 & 0.000785 \\
TESS (this study)& 0.117233 $\pm$ 0.000454 & 2459152.200533 $\pm$ 0.000154 & 88.415422 $\pm$ 0.412857 & 0.001749 \\
TESS (this study)& 0.117431 $\pm$ 0.000480 & 2459165.800856 $\pm$ 0.000153 & 88.291654 $\pm$ 0.336392 & 0.001731 \\
TESS (this study)& 0.117304 $\pm$ 0.000444 & 2459159.000638 $\pm$ 0.000143 & 88.147033 $\pm$ 0.260625 & 0.001661 \\
TESS (this study)& 0.116017 $\pm$ 0.000440 & 2459149.480297 $\pm$ 0.000159 & 88.157882 $\pm$ 0.301488 & 0.001661 \\
\enddata
\tablecomments{TESS observations in this table were acquired from the Mikulski Archive for Space Telescopes (MAST).   }
\end{deluxetable}

\subsection{Eclipse Data} \label{sec:eclipsedata}
The derivation of accurate and precise eclipse ephemerides is also critical for the efficient use of space-based observatory time \citep[e.g.,][]{zellem2020,pearson2022,burtzellem2024}. We used six mid-eclipse times to derive the mid-eclipse ephemeris for this study (Table~\ref{table:eclipsedata}). Eclipse data for WASP-77 A b came from JWST NIRSpec \citep[Program GTO 1274][]{august2023} with NRS1 and NRS2  covering the wavelength ranges of 2.674 $-$ 3.716 $\mu$m and 3.827 $-$ 5.173 $\mu$m, respectively. We utilized eclipse light curves from HST/WFC3+G141 grism between 1.1 and 1.7$\mu$m \citep[Program GO-16168][]{mansfield2022}. Two of the six mid-eclipse times were estimated from publicly-available phase curves obtained by the Spitzer Space Telescope \citep{werner04} IRAC2 \citep{fazio2004infrared} at 3.6 and 4.5~$\mu$m (Program GO-13038).  For more information about the analyses of the eclipse data, see \nameref{sec:eclipsereudction} Section \ref{sec:eclipsereudction}.  Two additional mid-eclipse times were taken directly from Spitzer (Program GO-10102) and derived by \cite{garhart2020}.

\begin{deluxetable}{l l l}[H]
\tablecaption{Mid-Eclipse Times Used In This Study\label{tab:mideclipsedata}}
\tablehead{
\colhead{Source} & \colhead{Mid-eclipse (BJDTDB)} & \colhead{Mid-eclipse Uncertainty (days)}
}
\startdata  
JWST (this study)& 2459816.57496 & 0.00013 \\
HST (this study)& 2459161.04312 & 0.0003 \\
Spitzer IRAC2 3.6$\mu$m (this study)& 2457854.05090 & 0.00043 \\
Spitzer IRAC2 4.5$\mu$m (this study)& 2457722.12849 & 0.00041 \\
Spitzer IRAC2 3.6$\mu$m \cite{garhart2020} &2456975.47418 & 0.00070 \\
Spitzer IRAC2  4.5$\mu$m \cite{garhart2020} &2456978.19514 & 0.00076\\
\enddata
\tablecomments{This table lists the six mid-eclipse times that were used in joint fit analysis in \nameref{subsec:jointfit} Section \ref{subsec:jointfit}. }
\label{table:eclipsedata}
\end{deluxetable}

\subsection{Radial Velocity Data} \label{sec:rvdata}
We considered radial velocity observations from five public sources, ultimately selecting three for inclusion in the final analysis. Two data sets came from the WASP-77 A b discovery paper \citep[][Table \ref{tab:rvdiscovery}]{maxted2013}. The first (Program 088.C-0011(A)), from the HARPS spectrograph on the 3.6-meter telescope at the European Southern Observatory's (ESO) La Silla Observatory in Chile \citep{mayor2003}. The second was from CORALIE, a high-resolution radial velocity spectrograph which is attached to the 1.2-meter telescope at La Silla \citep{queloz2000}. We also included a set of HARPS radial velocities (Program 0102.C-0618(A)) taken after the instrument's fiber upgrade in 2015. These post-upgrade HARPS data (hereafter referred to as HARPS15) are obtained from the December 11, 2023 ESO/HARPS Radial Velocities Catalog (\citep{barbieri2023eso} see Table \ref{tab:rvharps}). 

We identified an additional set of radial velocity observations taken with the SOPHIE spectrograph, located at the Haute-Provence Observatory in southern France \citep{bouchy2009}. The SOHPIE data could not be confirmed by the SOPHIE team as having passed requisite quality checks, thus, these data were excluded from our analysis. We also identified data taken with ESO's radial velocity spectrograph, ESPRESSO \citep{pepe2021}, but noted that this data was taken during the planet's transit as part of a Rossiter-McLaughlin \citep{rossiter1924detection,mclaughlin1924some} analysis, and thus did not include these data in our analysis.  To accurately detect the influence of WASP-77 A b on the host star's orbital velocity, individual points taken during Rossiter-Mclaghlin effect were excluded which can contain shifts in the measured radial velocity.  One radial velocity measurement was removed from CORALIE and eight from HARPS15 that were taken during transit.  Our final data set for analysis included 8 radial velocity measurements from HARPS, 10 from CORALIE, and 14 from HARPS15.

\begin{deluxetable}{llll}[H]
\tabletypesize{\scriptsize}
\centering
\tablecaption{WASP-77~A~b Archival Radial Velocity Measurements from HARPS and CORALIE Spectrographs from Discovery\label{tab:rvdiscovery}}
\tablehead{
\colhead{Instrument} & \colhead{BJDTBD} & \colhead{RV (m/s)} & \colhead{RV uncertainty (m/s)}
}
\startdata
HARPS  &  2455832.8615  &  1462.6  &  2.0  \\
HARPS  &  2455832.9040  &  1511.1  &  2.4  \\
HARPS  &  2455832.9110  &  1523.3  &  2.5  \\
HARPS  &  2455889.7458  &  1406.3  &  3.6  \\
HARPS  &  2455890.5370  &  2022.0  &  3.7  \\
HARPS  &  2455890.7386  &  1865.6  &  4.1  \\
HARPS  &  2455891.5721  &  1767.0  &  4.4  \\
HARPS  &  2455891.7468  &  1988.5  &  4.6  \\
CORALIE  &  2455069.7803  &  1324.6  &  4.9  \\
CORALIE  &  2455163.6091  &  1336.2  &  5.2  \\
CORALIE  &  2455169.6920  &  1967.3  &  4.9  \\
CORALIE  &  2455170.6625  &  1570.6  &  5.3  \\
CORALIE  &  2455188.6258  &  1969.6  &  5.4  \\
CORALIE  &  2455856.7095  &  1865.5  &  5.3  \\
CORALIE  &  2455914.6783  &  1710.5  &  4.3  \\
CORALIE  &  2455915.6775  &  1332.8  &  4.4  \\
CORALIE  &  2455916.6681  &  1708.7  &  4.8  \\
CORALIE  &  2455917.6500  &  1954.4  &  4.7  \\
CORALIE  &  2455918.6645  &  1554.2  &  6.5  \\
\enddata
\tablecomments{The table contains the radial velocity (RV) data from the discovery paper \citep{maxted2013} used in the radial velocity analysis as described in \nameref{subsec:rva} Section \ref{subsec:rva}.  Radial velocity measurement from 2455916.6681 BJDTDB is removed from analysis as it was taken during transit with Rositter-McLaughlin effect.}
\end{deluxetable}

\begin{deluxetable}{llll}[H]
\tabletypesize{\scriptsize}
\centering
\tablecaption{WASP-77~A~b Archival Radial Velocity Measurements from HARPS Post 2015 Spectrograph Upgrade\label{tab:rvharps}}
\tablehead{
\colhead{Instrument} & \colhead{BJDTBD} & \colhead{RV (m/s)} & \colhead{RV uncertainty (m/s)}
}
\startdata
HARPS15 & 2458427.787554 & 1470.018000 & 1.573475 \\
HARPS15 & 2458428.560896 & 1879.973206 & 1.785650 \\
HARPS15 & 2458428.570433 & 1868.396660 & 2.194125 \\
HARPS15 & 2458428.582250 & 1851.242091 & 2.868588 \\
HARPS15 & 2458428.593025 & 1831.094489 & 2.098265 \\
HARPS15 & 2458428.603905 & 1817.661492 & 2.013355 \\
HARPS15 & 2458428.614680 & 1806.211830 & 1.762600 \\
HARPS15 & 2458428.625664 & 1807.350545 & 1.740453 \\
HARPS15 & 2458428.636231 & 1812.432497 & 1.752132 \\
HARPS15 & 2458428.646902 & 1783.755312 & 1.610181 \\
HARPS15 & 2458428.657677 & 1750.562875 & 1.630013 \\
HARPS15 & 2458428.668452 & 1713.482612 & 1.643702 \\
HARPS15 & 2458428.679135 & 1680.066842 & 1.828763 \\
HARPS15 & 2458428.690003 & 1649.613223 & 1.723875 \\
HARPS15 & 2458428.700778 & 1647.376523 & 1.634059 \\
HARPS15 & 2458428.711762 & 1653.280585 & 1.612155 \\
HARPS15 & 2458428.722444 & 1638.297547 & 1.559323 \\
HARPS15 & 2458428.733219 & 1624.977366 & 1.527627 \\
HARPS15 & 2458428.743891 & 1608.904283 & 1.506273 \\
HARPS15 & 2458428.754562 & 1595.072828 & 1.538156 \\
HARPS15 & 2458428.765441 & 1578.391821 & 1.595225 \\
HARPS15 & 2458428.776216 & 1.568452064 & 1.563796 \\
\enddata
\tablecomments{Radial velocity (RV) data from the December 11, 2023 ESO/HARPS Radial Velocities Catalog (\citep{barbieri2023eso} used in this study. Eight measurements taken during the planets transit that were affected by Rositter-Mclaughlin effect, are removed from analysis for measurements between 2458428.63 and 2458428.71 BJDTDB. The remaining 14 radial velocity observations listed here are fit as described in \nameref{subsec:rva} Section \ref{subsec:rva}.}
\end{deluxetable}

\section{Analysis} \label{sec:anal}

Using these transit photometry, eclipse photometry and radial velocity data, we performed a joint orbital analysis for WASP 77 A b. Our methodology adheres to the process described in \cite{pearson2022}.  

\subsection{Transit Photometry Reduction} \label{sec:transitreduction}

We modeled individual light curves for each photometric time series to derive the best-fit parameters including transit depth, inclination, and mid-transit time. We performed model fitting of transit light curves using EXOTIC (EXOplanet Transit Interpretation Code), a comprehensive Python data reduction tool designed to support the citizen science and professional community in analyzing exoplanet transit data \citep{zellem2020,Fatahi2024prep}. The code uses priors from the NASA Exoplanet Archive, ensuring the use of the most-recently published system parameters. EXOTIC yields full posteriors and uncertainties for each parameter.

To enhance the precision of our parameter estimates, we employed UltraNest \citep{ultranest2016}, a robust multimodal nested sampling algorithm. To prevent the influence of data points that might not represent significant transit signals, light curves with residuals larger than the transit depth were removed to ensure that the model's fit captured significant transit signals from WASP-77~A~b. 
Individual data points failing a 3$\sigma$ clip were discarded prior to fitting the light curve. Observations with a flux standard deviation greater than 0.03 were also excluded to eliminate data with excessive noise, as high scatter can indicate instability in the observations or environmental contamination. Observations with airmass values larger than 2 were excluded.  

For ExoClock and ETD observations that did not include airmass or the elevation of the observatory, we calculated the airmass values by extracting the latitude and longitude of the observer's location and then used the Google Maps API to derive the elevation based on geographic data. This geographic information was then passed to EXOTIC to calculate the airmass values for each time stamp of the corresponding observation.

Limb darkening coefficients depend on the host stellar parameters and the wavelength at which the observation was taken and must be derived to fit the model accurately. 
Based on the filter used for the observations, we assigned a corresponding passband within EXOTIC utilizing the Python package \texttt{Exotethys} \citep{Morello2020} with stellar parameters for
effective temperature 5617 $\pm7.2\times10^{1}$ kelvin,
Metallicity (dex) -0.10 $\pm1.1\times10^{-1}$ [Fe/H],
and stellar surface gravity 4.476 $\pm1.5\times10^{-2}$ $\log g \left( \log_{10}(cm/s^{2}) \right)$ \citep{cortes2020tramos} (see Table \ref{tab:stellarparameterpriors}) 
to derive the limb darkening coefficients for each fit.

The citizen science mid-transit times and TESS data were incorporated into a comprehensive joint transit, eclipse and radial velocity analysis described in \nameref{subsec:jointfit} Section \ref{subsec:jointfit}. (Refer to Tables \ref{tab:Exoplanet Watch Single Fit Outputs}, \ref{tab:exoclocktmids}, \ref{tab:etdtmids} and \ref{tab:TESS Fits} for the list of mid-transit times from Exoplanet Watch, ExoClock, ETD, and TESS.)  Properly model fitting the individual transit data accounting for airmass and limb darkening effects provided constraints on the transits' inclination, RpRs, and most relevant to our study, precise mid-transit times for our joint transit, eclipse and radial velocity analysis.  The resulting citizen science individual mid-transit times were used as priors in our joint fit analysis in \ref{subsec:jointfit}.

\begin{deluxetable}{| p{2.8cm} |p{2.6cm} | p{2.4cm} |}
\tabletypesize{\scriptsize}
\tablecaption{Stellar Priors Used for This Study 
\label{tab:stellarparameterpriors}}
\tablehead{
\colhead{Parameter} & 
\colhead{Units} & \colhead{\citep{cortes2020tramos}}   
}
\startdata
$M_* \;$   
&  mass $[M_{\odot}]$
& $0.903 \pm 6.6\times10^{-2}$\\
\hline
$R_* \;$  
& radius $[R_{\odot}]$
& $0.910 \pm 2.5\times10^{-2}$ \\
\hline
$T_{\rm eff} \;$   
& Effective temperature $[{\rm K}]$
&  $5617 \pm 7.2\times10^{1}$\\
\hline
Metallicity (dex)   
& $[{\rm Fe/H}]$
&  $-0.10 \pm 1.1\times10^{-1}$\\
\hline
Stellar surface gravity  
& $\log g \left( \log_{10}({\rm cm/s^{2}}) \right)$
&  $4.476 \pm 1.5\times10^{-2}$\\
\hline
\enddata 
%\tablecomments{}
\end{deluxetable}

\subsection{Eclipse Data Reduction} \label{sec:eclipsereudction}

The two JWST eclipse light curves from \cite{august2023} were simultaneously fit together here to derive one mid-eclipse time via the same global fit process used for the transit data in \nameref{sec:globaltransitfit} Section \ref{sec:globaltransitfit}. The same process is used for two HST eclipse light curves from \cite{mansfield2022}.
The Spitzer eclipse data were processed using a public pipeline based on the pixel-map method \citep{Lewis2013} with additional optimizations for handling the ramp effect and uncertainty estimation using nested sampling. The pipeline uses gaussian kernel regression with a nearest neighbor technique to estimate the intrapixel sensitivities in each dataset. Non-physical solutions can arise if the priors are too large and not correlated to one another with consequences manifesting as negative night-side temperatures and wildly varying hot-spot offsets compared to GCM models. To avoid non-physical solutions, we used conditional priors to model the phase curve components (i.e., day-night amplitude and hot-spot offset) which is important for deriving physically-plausible solutions when using a linear system of equations. Two additional Spitzer mid-eclipse (emid) values were adopted from \cite{garhart2020}.

%%%%%%%%%%%%%%%%%%%%%%%%%%%%%%%%%%%%%%%%%%%%%
%%       Transit-Only Fit SubSection       %%
%%%%%%%%%%%%%%%%%%%%%%%%%%%%%%%%%%%%%%%%%%%%%
\subsection{Global Fits of Citizen Science and TESS Transit Data for Comparison} \label{sec:globaltransitfit}
The lightcurves individually fit for better constraints on transit parameters described in \nameref{sec:transitreduction} Section \ref{sec:transitreduction}, were combined and fit simultaneously here in a global fit in order to compare the use of citizen science and professional transit data sets. The inclination, planet-to-star radius ratio, and the mid-transit time were left as free parameters in the individual fits and, once derived, were passed to the global fit for more precise constraints on these parameters with the addition of the orbital period. For the global fit, in order to ensure the ephemeris remains fresh longer, we adopted the mid-transit time from the most recent available light curve in the corresponding data set as our weighted prior for the new ephemeris. See \nameref{sec:results} Section \ref{sec:results} for results and comparison of the global fits.

%%%%%%%%%%%%%%%%%%%%%%%%%%%%%%%%%%%%%%%%
%%       RV-only Fit SubSection       %%
%%%%%%%%%%%%%%%%%%%%%%%%%%%%%%%%%%%%%%%%
\subsection{Radial Velocity Analysis} \label{subsec:rva}
To determine if stellar variability impacted in the HARPS and CORALIE data, we searched for correlations between various activity indicators and the radial velocity data \citep{saar1997activity,desort2007search,lovis2011}. Specifically, we made use of two spectral line indicators (H-alpha and the S-index) and three metrics from the radial velocity Cross Correlation Functions (CCF), the FWHM, BIS, and Contrast. These values provide insights on whether the magnetic phenomena exhibited by a star, such as spots, plages, and granulation, are impacting the stellar absorption lines significantly enough to impact the measured radial velocities.

\begin{figure}[H]
\centering
\includegraphics[width=\textwidth]{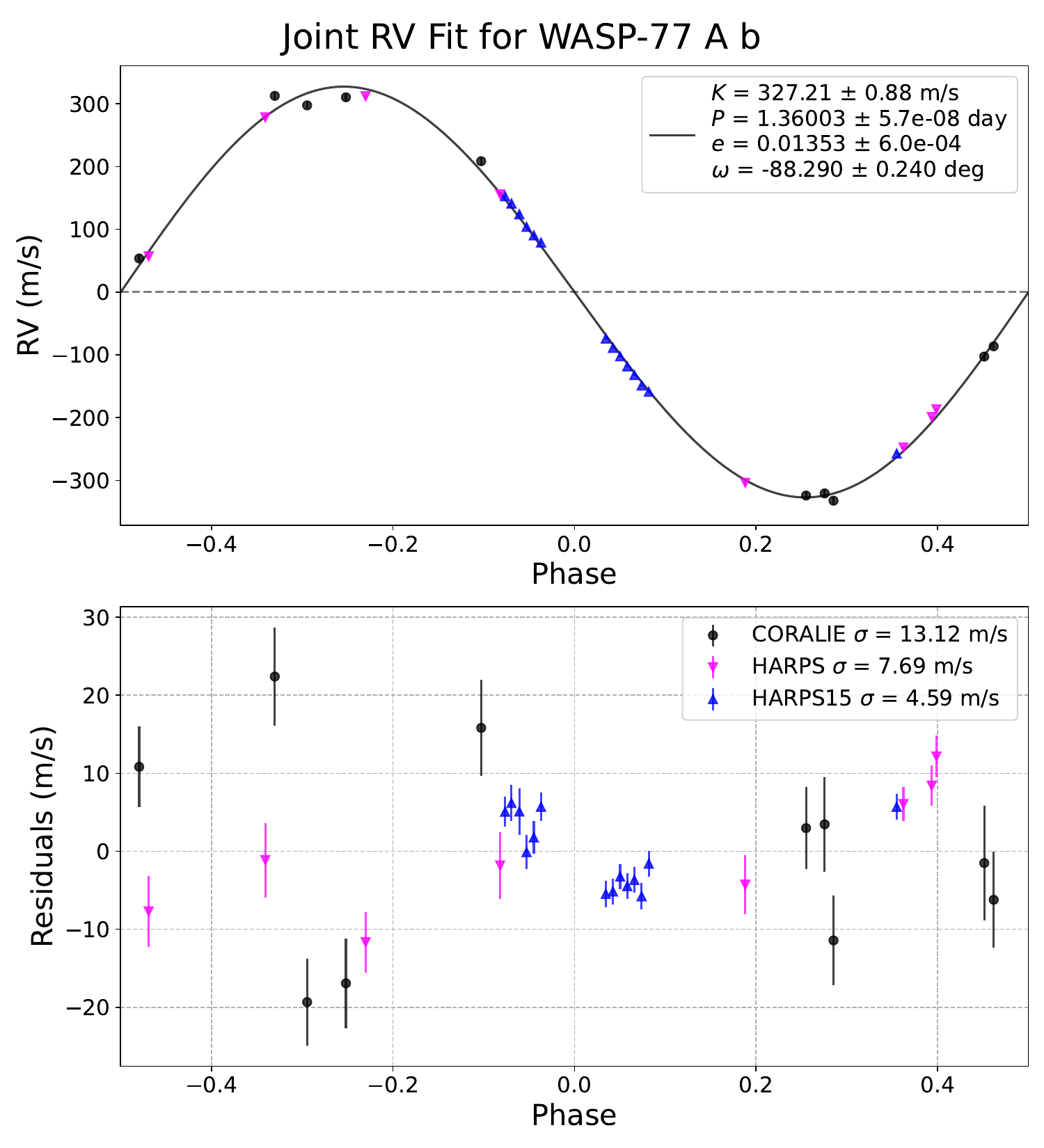}
\caption{Top: phase-folded radial velocity curve for WASP-77 A b. The radial velocity data included are from the discovery paper \cite{maxted2013} and recent HARPS data from the ESO archive. Bottom: radial velocity residuals for WASP-77 A b after removing the contribution for the planetary orbit.}
\label{fig:jointrv}
\end{figure}

After the individual indices were calculated, we measured their Pearson correlation coefficients with the radial velocity data. The highest-ranked index was the FWHM, which had a correlation coefficient of 0.58 with the radial velocities. As this was below the r = 0.6 limit generally used to identify moderate correlation \citep{akoglu2018userCoeffRV}, we asserted that WASP-77 A exhibits no significant stellar activity, and therefore we did not include activity-mitigation steps in our radial velocity fitting methodology. 
Radial velocity measurements spanning the full orbital phase of an exoplanet are crucial for determining accurate system parameters\citep{cumming1999lick, ford2005quantifying,hara2017radial}.   In particular, the shape of the orbit cannot usually be constrained by transit photometry alone \citep{kipping2008transiting}. By including radial velocities throughout the planet's orbit in our analysis Fig.~\ref{fig:jointrv}, we can determine the eccentricity of WASP-77 A b’s orbit accurately \citep{cumming1999lick, wright2009efficient}.
We adopted the \cite{pearson2022} methodology of using a Keplerian model to fit the radial velocity data and determine the orbital period, argument of periastron, eccentricity, and semi-amplitude of WASP-77~A~b. 

%%%%%%%%%%%%%%%%%%%%%%%%%%%%%%%%%%%%%%
%%       Joint Fit SubSection       %%
%%%%%%%%%%%%%%%%%%%%%%%%%%%%%%%%%%%%%%
\subsection{Joint Simultaneous Fit of Transit, Eclipse, and Radial Velocity Data} \label{subsec:jointfit}
We leveraged a joint-simultaneous fit of the transit, eclipse, and radial velocity data to place strong constraints on the system parameters, including new transit and eclipse ephemerides, following the prescription of \citet{pearson2022}. To do so, we coupled the underlying orbit equations between the planet and star to ensure a self consistent solution --- something that is not possible when using separate code-bases for each respective measurement type. We updated WASP-77~A~b’s parameters for orbital period, mid-transit time, inclination, argument of periastron, eccentricity, planet mass, and the radius ratio between the planet and the star with a nested sampler. Our joint analysis utilized the stellar priors in Table~\ref{tab:stellarparameterpriors} from \cite{cortes2020tramos}. The joint optimization leveraged a combined likelihood function with contributions from radial velocity, photometry, and ephemerides: $L_{joint} = L_{RV} + L_{transit}+L_{mid-transit}+L_{mid-eclipse}$. 

\begin{figure}[H]
\includegraphics[width=0.98\textwidth]{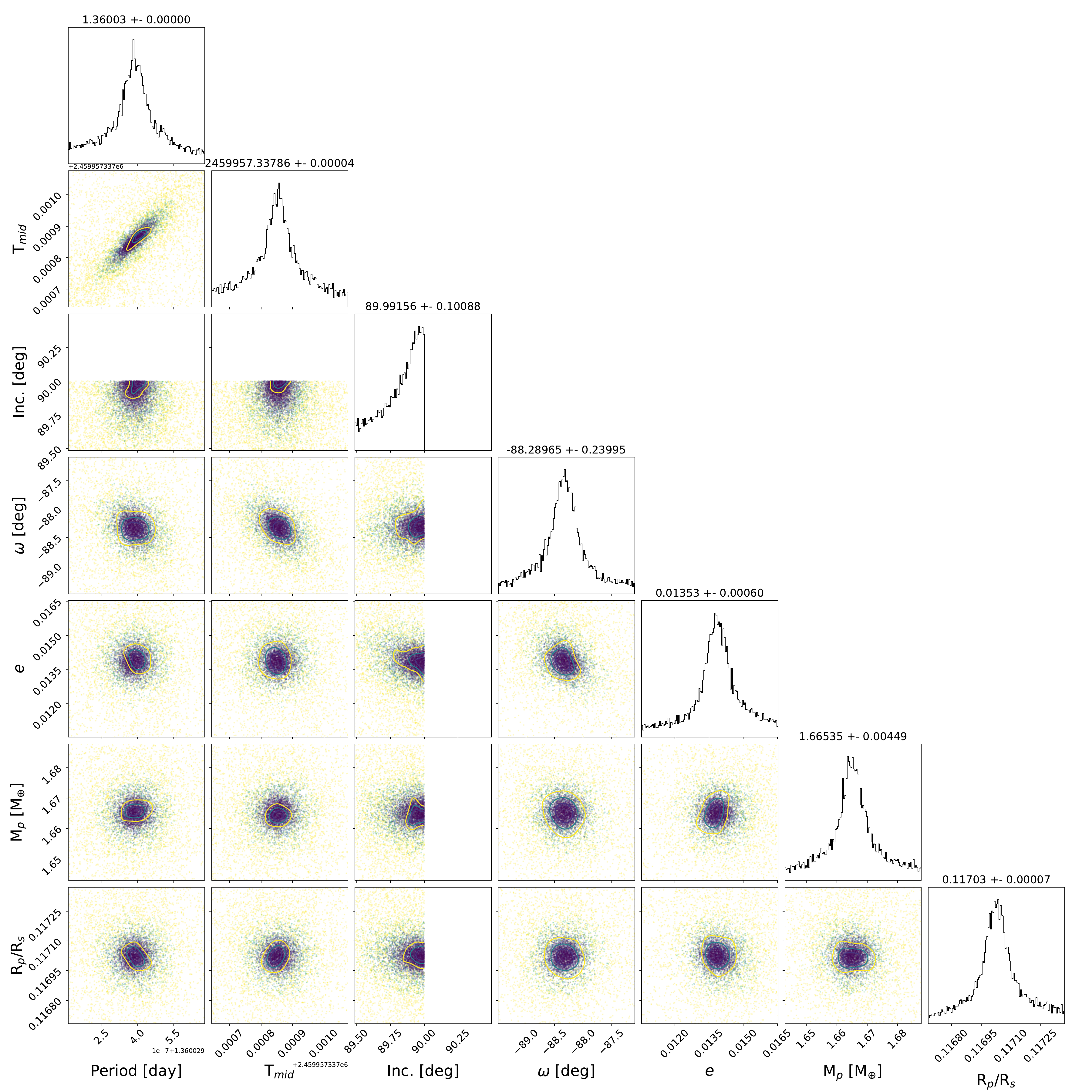}
\caption{
Posterior plot depicting the most probable parameter estimates and their uncertainties from our joint transit, eclipse and radial velocity data fit for period, mid-transit time, inclination, argument of periastron, eccentricity, mass of planet, and planet star radius ratio. The contour plots indicate the density of the probability distribution with the inner contour representing a 1 sigma deviation. The shape and orientation of the contour plots are indicative of the amount of correlation between parameters with a more circular shape depicting less correlation. 
The narrow peaks in the figure represent the most probable estimates with a Gaussian distribution illustrating uncertainties for fit parameters. In the case of inclination, we deliberately restrict the upper limit to 90° to reflect the physical limitation on this parameter; consequently, the cone is truncated on right side for inclination.}
\label{fig:jointposteriors}
\end{figure}

The final posterior distributions from the nested sampling algorithm can be seen in Figure \ref{fig:jointposteriors} depicting the most probable parameter estimates and their uncertainties.  The Gaussian distribution about the narrow likelihood maximum peak ensures we effectively explore the parameter space and that our estimates represent the data.

\begin{figure}[H]
\includegraphics[width=1.0\textwidth]{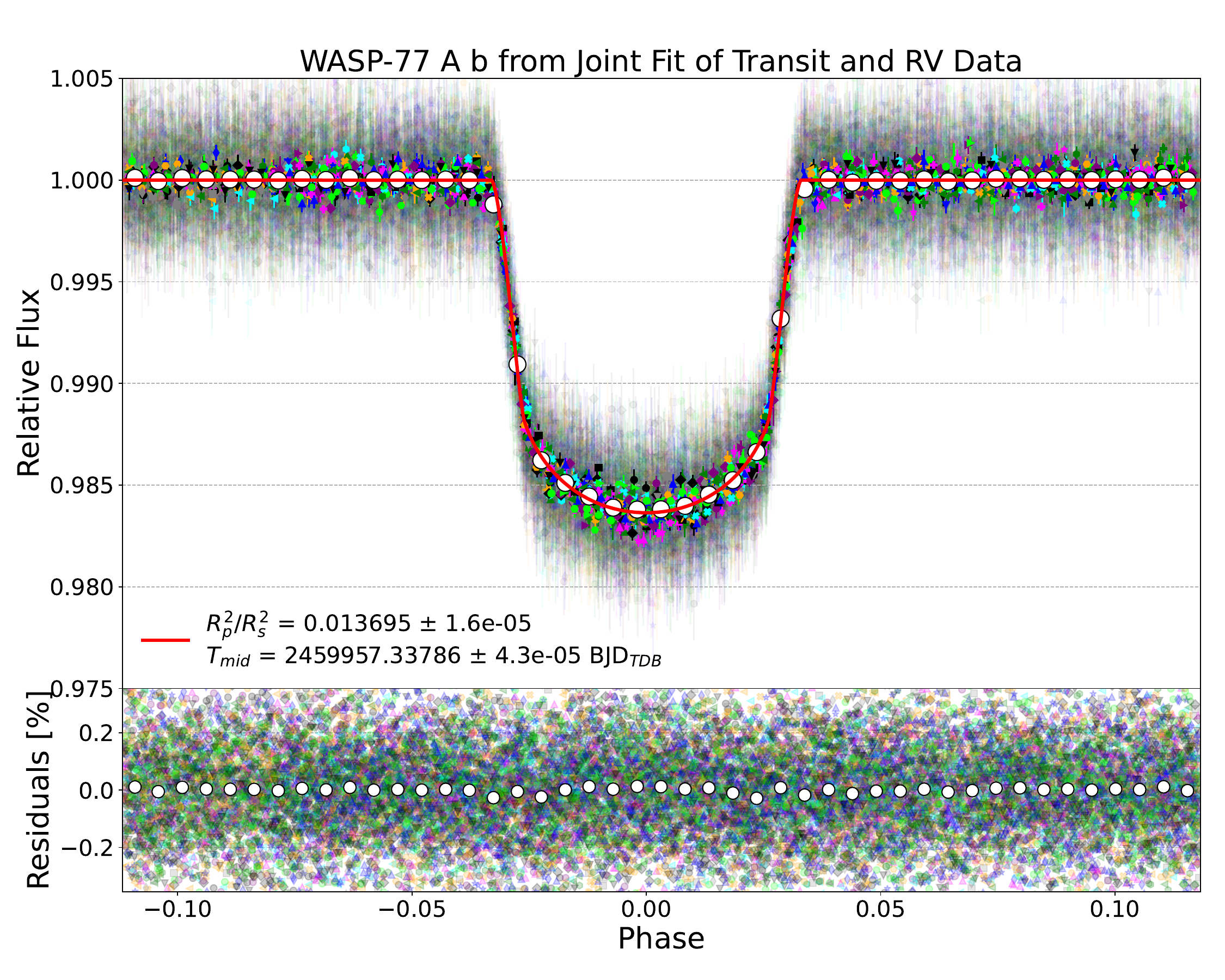}
\caption{
We present the phase-folded TESS transit data from the joint simultaneous fit of WASP-77~A~b.  The top plot is a stacked light curve plot from \nameref{subsec:jointfit} Section \ref{subsec:jointfit} depicting the wellness of fit. The transit data is displayed as colored points, with the predicted light curve displayed as a red line. The lower plot is the residuals illustrating the difference between the observation data and the model.} 
\label{fig:jointglobalplot}
\end{figure}

\begin{figure}[H]
\includegraphics[width=.85\textwidth]{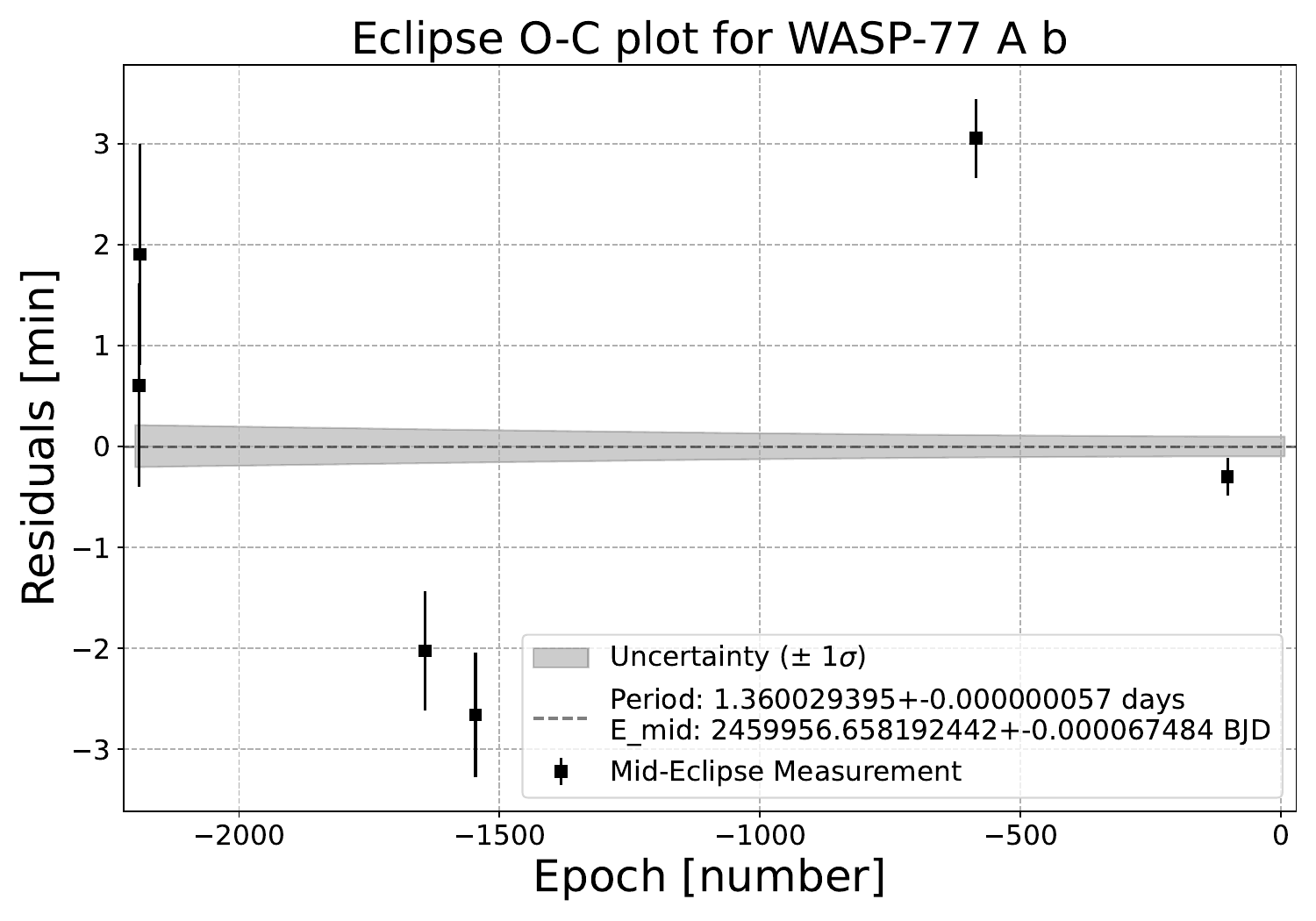}
\caption{Eclipse O-C plot for WASP-77~A~b with mid-eclipse times as listed in Table \ref{tab:mideclipsedata}. The residuals are shown in minutes and represent the difference between the observed and calculated mid-eclipse times based on a fixed linear ephemeris. The grey bar denotes a $1\sigma$ uncertainty. This constricted grey region illustrates the power of our analysis in ensuring a fresh eclipse ephemeris.} 
\label{fig:eclipseoc}
\end{figure}

\begin{figure}[H]
\includegraphics[width=0.84\textwidth]{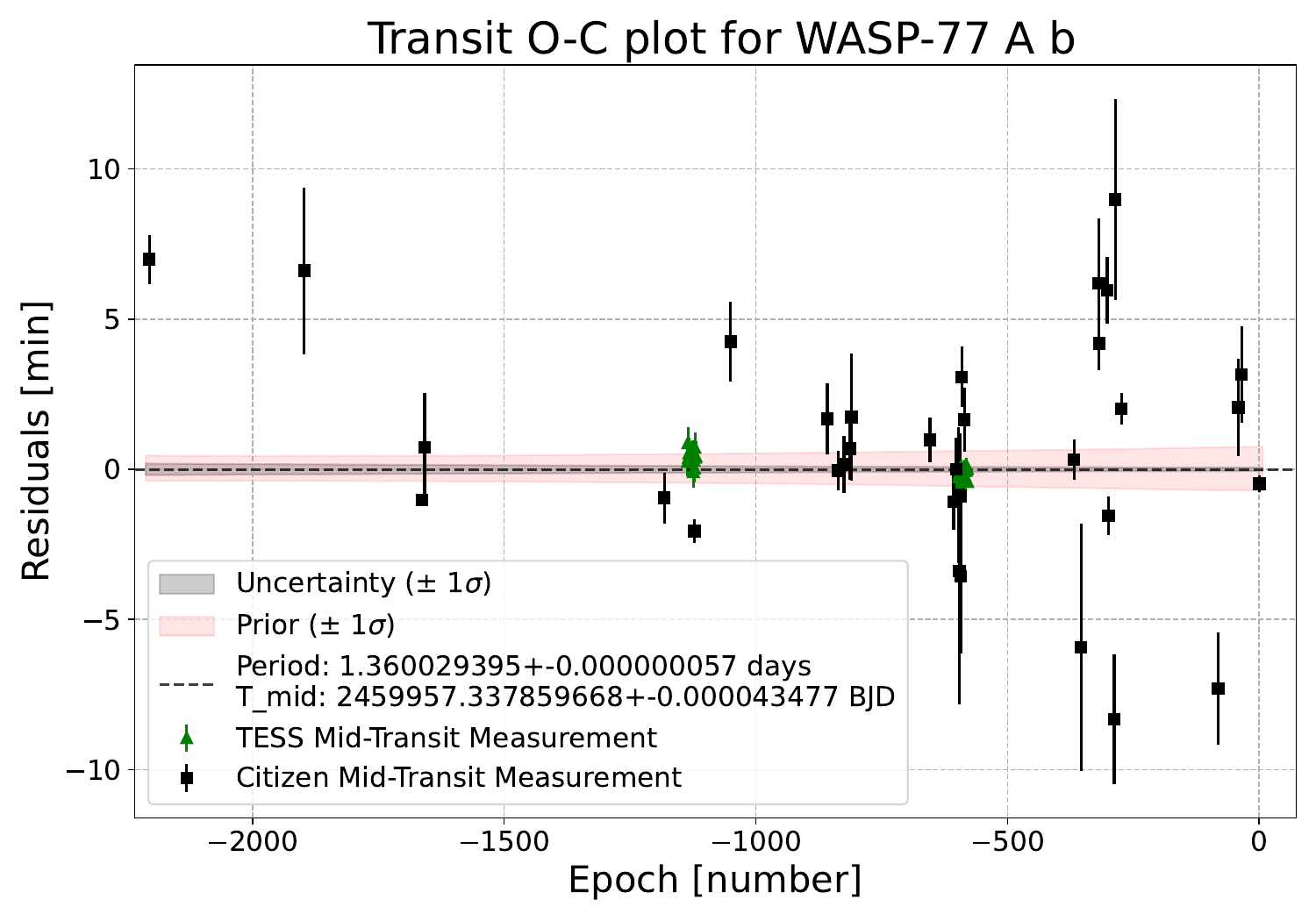}
\caption{O-C plot for mid-transit times of WASP-77~A~b based on our joint analysis. The residuals are shown in minutes and represent the difference between the observed and calculated mid-transit times based on a fixed linear ephemeris. The grey bar denotes a $1\sigma$ uncertainty. The pink shaded area indicates the propagated uncertainty of the prior ephemeris without incorporating the updates from our current study. The distribution of residuals around the zero line validates the precision of the updated orbital parameters.}
\label{fig:jointocplot}
\end{figure}

The stacked light curves in Figure \ref{fig:jointglobalplot} from TESS, display the wellness of fit when analyzed simultaneously with eclipse and radial velocity data in the joint analysis. The TESS stacked lightcurves yield a a 856$\sigma$ detection of the transit depth ${R_p/R_*}^2$ 0.013695 $\pm1.6\times10^{-5}$. Mid-transit times for the citizen science data were incorporated into the joint fit as priors. For individual fit plots of the transit data, see Figures \ref{fig:citiindividualplots} and \ref{fig:tessindividualplots}. Table \ref{tab:jointfitparameters} presents our joint fit final parameters with comparison to prior work described in \nameref{sec:results} Section \ref{sec:results}.  The observed minus calculated (O-C) diagram presented in Figure \ref{fig:eclipseoc} illustrates the mid-eclipse timing residuals for WASP-77~A~b from our joint analysis. The grey error bars represent a $1\sigma$ uncertainty from the predicted mid-eclipse time. We also produce an O-C plot for WASP-77~A~b's mid-transit times (Fig.~\ref{fig:jointocplot}) which indicates how much our ephemeris derived here (grey-shaded region) improves upon the current ephemeris uncertainty in the published literature \citep{cortes2020tramos} (pink shaded region).

\begin{deluxetable}{|p{1.4cm}|p{2.0cm}|p{2.4cm}|p{3cm}|p{2.0cm}|p{1.8cm}|}
\tabletypesize{\scriptsize}
\tablecaption{System Parameters for WASP-77 A b Joint Fit Comparison with Prior Papers
\label{tab:jointfitparameters}}
\tablehead{
\colhead{Parameter} & 
\colhead{Units} & \colhead{\citep{cortes2020tramos}} & \colhead{\citep{kokori2023}} & \colhead{This Work}  
}
\startdata
\hline
$P$ 
& Orbital Period [days]
& $1.36002854 \pm 6.2\times10^{-7}$ 
& $1.360028949 \pm 7.5\times10^{-8}$ 
& $1.360029395 \pm 5.7\times10^{-8}$ \\
\hline 
$T_{mid}$ 
& Mid-transit time [BJDTDB] 
& $2457420.88439 \pm 8.5\times10^{-4}$ 
& $2458693.870688 \pm 3.9\times10^{-5}$ 
& $2459957.337860 \pm 4.3\times10^{-5}$ \\
\hline
$E_{mid}$  
& Mid-Eclipse time [BJDTDB] 
& $2457658.2054 \pm 4.4\times10^{-3}$
& $n/a$
& $2459956.658192 \pm 6.7\times10^{-5}$\\
\hline
$i$  
& Inclination [deg]
& $88.91 \pm 9.5\times10^{-1}$ 
& $n/a$
& $89.99 \pm 1.0\times10^{-1}$\\
\hline
$\omega$  
& Argument of periastron [deg]
& $-166 \pm 7.5\times10^{+1}$ 
& $n/a$
& $-88.29 \pm 2.4\times10^{-1}$ \\
\hline
$e$  
& Eccentricity
& $0.00740 \pm 6.9\times10^{-3}$ 
& $n/a$
& $0.01353 \pm 6.0\times10^{-4}$\\
\hline
$M_p$ 
& Mass of planet [$M_{Jup}$] 
& $1.667 \pm 6.8\times10^{-2}$ 
& $n/a$
& $1.6654 \pm 4.5\times10^{-3}$\\
\hline
$R_p/R_*$  
& Radius of planet in stellar radii
& $0.13354 \pm 7.4\times10^{-4}$ 
& $n/a$
& $0.117026 \pm 6.9\times10^{-5}$\\
\hline
$a$ 
& Semi-major axis [AU]
& $0.02335 \pm 4.5\times10^{-4}$
& $n/a$
& $0.023233633 \pm 3.7\times10^{-8}$\\
\hline
$a/R_*$  
& Semi-major axis in stellar radii
& $5.332 \pm 8.1\times10^{-2}$
& $n/a$
& $5.4900859 \pm 8.6\times10^{-6}$\\
\hline
$K$ 
& Radial Velocity semi-amplitude [m/s]
& $323.4 \pm 3.8\times10^{0}$
& $n/a$
& $327.21 \pm 8.8\times10^{-1}$\\
\hline
Propagated $T_{mid}$ to 2030-01-01 
& [BJDTDB] 
& $2462503.31104 \pm 2.47\times10^{-3}$ ($\pm 213$ sec) 
& $2462503.311774 \pm 2.14\times10^{-4}$ ($\pm 18$ sec) 
& $2462503.312887 \pm 1.15\times10^{-4}$ ($\pm 9.9$ sec) \\
\hline 
Propagated $E_{mid}$ to 2030-01-01 
& [BJDTDB] 
& $2462502.6271 \pm 4.9\times10^{-3}$ ($\pm 425.35$ sec)
& $n/a$
& $2462502.633219 \pm 1.26\times10^{-4}$ ($\pm 10.89$ sec) \\
\hline
\enddata 
\tablecomments{This table cross-references the parameters in the joint transit, eclipse, and radial velocity fit to prior publications on WASP-77 A b: \citep{cortes2020tramos,kokori2023}.}
\end{deluxetable}

\subsection{Transit Timing Variation Analysis} \label{subsec:ttv}
\begin{figure}[H]
\includegraphics[width=0.85\textwidth]{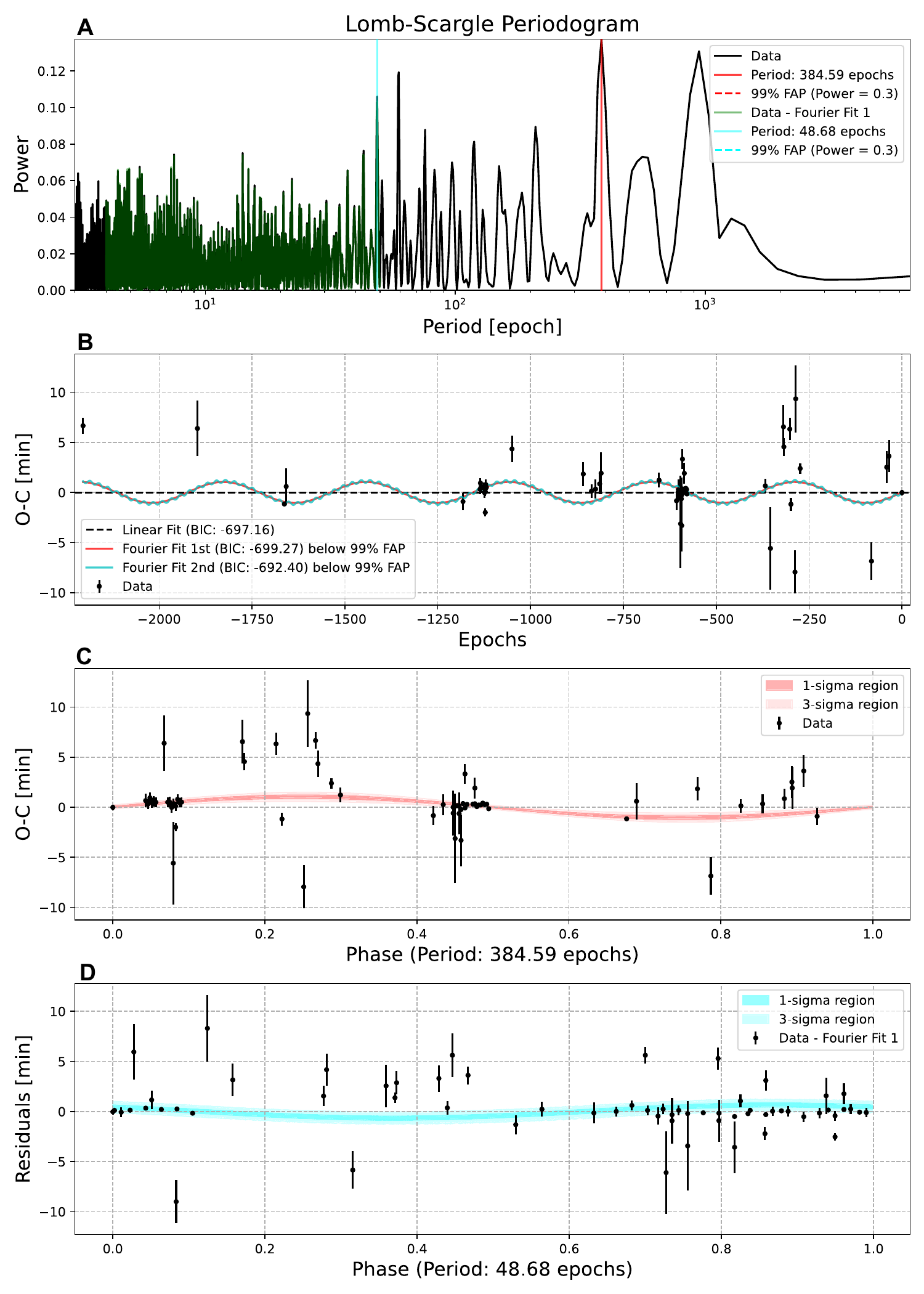}
\caption{Diagnostic plots to search for transit timing variations using a periodogram analysis with the Lomb-Scargle algorithm. A) Power versus orbital period plot indicating the potential for underlying periodic signals. “Detrended” data also presented. B) O-C data overlaid with single-period (Fourier Fit 1) and two-period (Fourier Fit 2) models. C) A phase-folded solution for Fourier fit 1. Based on the BIC value reported in B, we do not find a significant periodic trend with a single term. D) A phase-folded plot for the second term in Fourier fit 2 after subtracting the first term. For more details, please see TTV Analysis \nameref{subsec:ttv} Section \ref{subsec:ttv}.}
\label{fig:periodogram}
\end{figure}

 Transit timing variations (TTVs), periodic perturbations to the mid-transit times of a known planet, can indicate the presence of additional planets in a system \citep[e.g.,][]{agol2005detectingTTV,holman2005useTTV,ford2012transitTTV,hadden2016numericalTTV,pearson2019}. TTVs can reveal long-term periodic trends, usually associated with precession, as well as short-order perturbations \citep{ragozzine2010valueTTV,deck2014orbitalTTV,pearson2019,korth2020characterization}, so a two-term model is needed to capture the dynamics of both. We generated Lomb-Scargle periodograms to look for Transit Timing Variations (TTVs) in the O-C data residuals, indicating potential periodic signals in the power versus orbital period, Figure \ref{fig:periodogram}. 

For our WASP-77~A~b data, we calculated BIC values for a linear fit and a Fourier fit to the data to help us determine which model better represents the data. If there is periodicity in the mid-transit times, then the Fourier fit will have the lower BIC value as it can detect periodicity that cannot be detected by the linear model \citep{schwarz1978estimating}. When comparing Bayesian Information Criterion (BIC) values between different models, the model with the lowest BIC is typically considered the best fit to the data \citep{schwarz1978estimating} and is derived by the equation:
\begin{equation}
BIC = k \ln(n) - 2 \ln(\hat{L}),
\end{equation}
 where \(k\) is the number of parameters estimated by the model, \(n\) is the number of observations or sample size, and \(\hat{L}\) is the maximized value of the likelihood function of the model. The term \(k \cdot \ln(n)\) penalizes complexity to avoid overfitting, while \(-2 \cdot \ln(\hat{L})\) pertains to the goodness of fit.

The false alarm probability (FAP) is a statistical measure used to assess the likelihood that a perceived signal in the data could arise from random fluctuations alone rather than any actual periodic variation \citep{xie2014frequencyTTV}. FAP values are ultimately used, along with Bayesian evidence, to signify the presence or absence of a periodic variation in our transit times.

Panel A describes the power versus orbital period. Panel B overplots the O-C data with our two Fourier models, while Panels C and D depict phase-folded solutions for Fourier fit 1 and 2 respectively, illustrating how well they correspond with the observed data. In the phase folded plots, the x-axis is in phase relative to orbital epochs determined by the mid-transit point of the orbit and period, indicating that the mid-transit times are plotted against the phase of the planet's orbit. Each point represents the deviation of the observed transit time from a predicted time, based on the orbital period.
Based on the BIC values, we do not find a significant periodic trend with a single term. We do not see significant evidence for a two-term periodic model and report no transit timing variations in the mid-transit times from this study.

\section{Results and Discussion} \label{sec:results}
Our joint analysis of the combined radial velocity, eclipse, and transit data set, constrains the orbital parameters of WASP 77 A b to high precision (Table \ref{tab:jointfitparameters}). 
We present a new 
orbital period of 1.360029395 $\pm$5.7$\times$10$^{-8}$ days, 
mid-transit time of 2459957.337860 $\pm$4.3$\times$10$^{-5}$ BJDTDB, 
mid-eclipse time of 2459956.658192 $\pm$6.7$\times$10$^{-5}$ BJDTDB,
inclination 89.99 $\pm1.0\times10^{-1}$ [deg],
argument of periastron -88.29 $\pm2.4\times10^{-1}$ [deg], 
eccentricity 0.01353 $\pm6.0\times10^{-4}$,
mass of planet 1.6654 $\pm4.5\times10^{-3}$ $M_{Jup}$,
radius of planet in stellar radii 0.117026 $\pm6.9\times10^{-5}$ $R_p/R_*$,
semi-major axis 0.023233633 $\pm3.7\times10^{-8}$ [AU],
semi-major axis in stellar radii 5.4900859 $\pm8.6\times10^{-6}$ $a/R_*$,
and radial Velocity semi-amplitude 327.21 $\pm8.8\times10^{-1}$ $K$ [m/s]. 

\subsection{Forward Propagation of Transit and Eclipse} \label{subsec:forward}

To accurately predict future transit and eclipse times and their uncertainties for WASP-77 A b, we employed an analytical error propagation method for forward propagation using the notation of \cite{zellem2020}. 
The precise prediction of an exoplanet's next transit time ($T_{\text{mid}}$) is given through the equation:
\begin{equation}
T_{\text{mid}} = n_{\text{orbit}} \cdot P + T_0
\end{equation}
where $n_{\text{orbit}}$ represents the number of orbits until the future transit event, $P$ is the orbital period of the planet, and $T_0$ is the established mid-transit or mid-eclipse time.
The analytical derivation of the uncertainty in the mid-transit time ($\Delta T_{\text{mid}}$) incorporates the propagation of errors in the orbital period ($\Delta P$) and the starting mid-transit time ($\Delta T_0$), as:
\begin{equation}
\Delta T_{\text{mid}} = \sqrt{(n_{\text{orbit}}^2) \cdot (\Delta P^2) + (\Delta T_0^2)}.
\end{equation}

The prior mid-transit time we constrained and propagated was taken from the most recent transit data which was from an Exoplanet Watch user. Using the most recent observation reduces the uncertainties that older published mid-transit times may have.  

For our joint fit analysis results of the citizen science and TESS transit data, eclipse data, and radial velocity data, we find a January 1, 2030 (a reasonable date for the start of the Ariel mission science operations \citep{burtzellem2024}) propagated mid-transit uncertainty of $\pm$9.94 seconds. Similarly, from the joint fit analysis we find a propagated mid-eclipse time uncertainty of $\pm$10.89 seconds.  Our forward-propagation results demonstrate how we can obtain a lower uncertainty in the propagated mid-transit time with the addition of all of the latest available data. Compared to a previously published mid-transit time from a similar study by \cite{cortes2020tramos}, we demonstrate we can achieve a tighter constraint on the forward propagated mid-transit time by a factor of 21.5 by including the latest data.  When our new mid-eclipse time is forward propagated to January 1, 2030, we improve precision by a factor of 39.1 compared to the previous study's propagated mid-eclipse time \citep{cortes2020tramos}.  

We also compared forward propagated mid-transit times from the global fits of the citizen science data alone, the TESS data alone, and then with these two combined sources, to quantify the benefit of incorporating citizen science data when updating exoplanet ephemerides. The resulting mid-transit times and errors were also forward propagated to the date January 1, 2030. The TESS data alone produces a mid-transit time uncertainty on January 1, 2030 of $\pm40.35$ seconds, while the citizen science data alone produces a propagated mid-transit time uncertainty of $\pm22.55$ seconds (Table \ref{tab:globalfitcomparison}). The lower uncertainty, despite the larger per-measurement uncertainty, may be due to a longer time baseline of the citizen science data giving an advantage in constraining some parameters like period and mid-transit time. The longer baseline may explain why the derived parameters have smaller uncertainties in the citizen science data alone fit, by enabling us to effectively bin down the noise and constrain the ephemeris more precisely than with the TESS data.  We can compensate for the lower SNR measurements with more measurements over a longer baseline \citep{dragomir20}. While this is a minimal difference in the observational padding required, it nonetheless demonstrates how the broader baseline of the citizen science observations increases the mid-transit time precision beyond what is possible with the higher precision, but more temporally concentrated, TESS data alone. Figure \ref{fig:jointocplot} depicts the concentrated TESS data and temporal spread of the citizen science data illustrating how these observations can aid in filling in gaps between TESS sectors. Unsurprisingly, our global fit analysis demonstrates that we obtain the most precise constraints when we combine both the citizen science and TESS photometry data, which produces a January 1, 2030 mid-transit time uncertainty of $\pm17.37$ seconds, a factor of 2.3 improvement over the TESS data alone. 

\begin{figure}[H]
\includegraphics[width=1.0\textwidth]{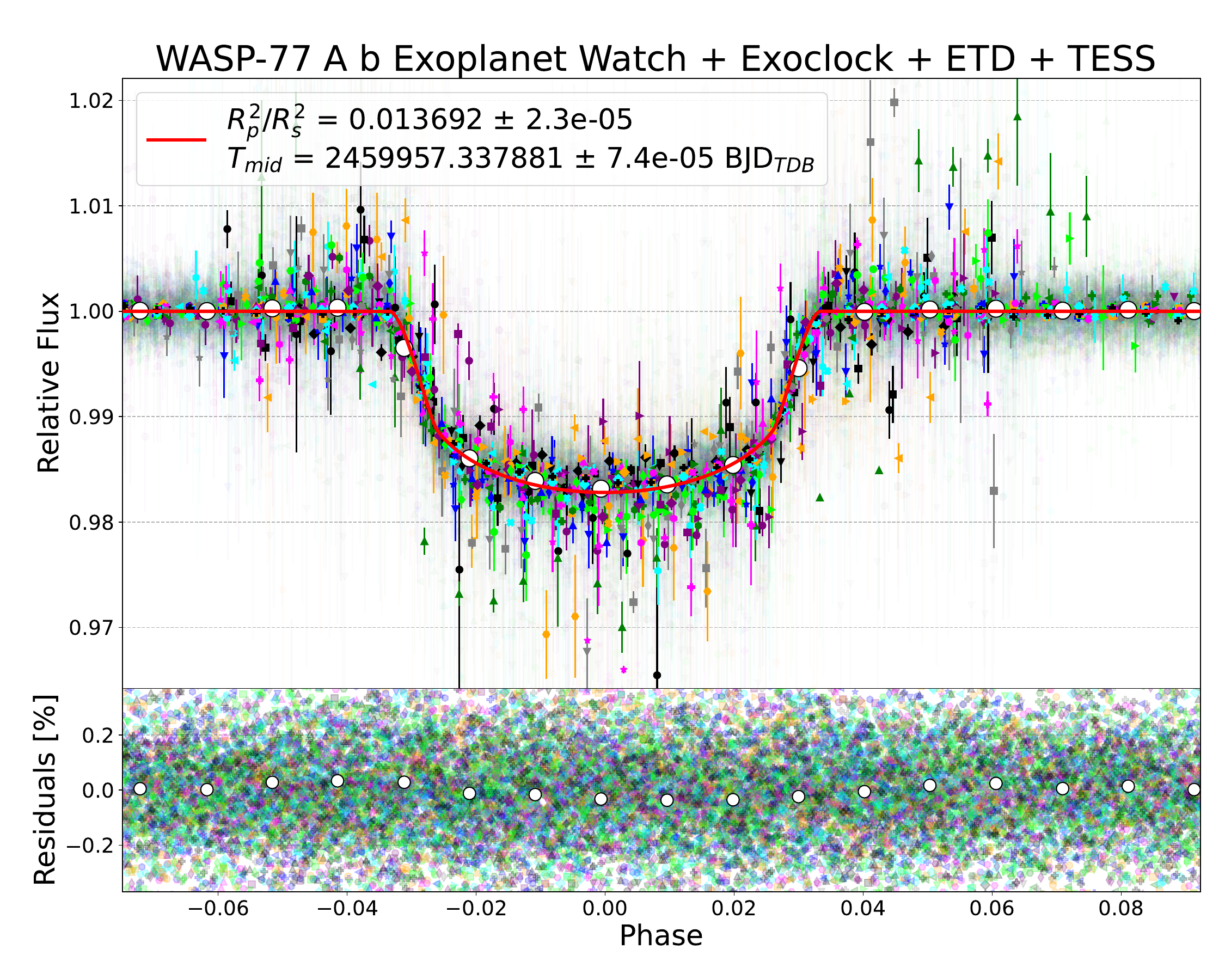}
\caption{
The top plot is a stacked light curve plot from global fit analysis described in \nameref{sec:globaltransitfit} Section \ref{sec:globaltransitfit} and \ref{subsec:forward} of WASP-77~A~b. The lower plots are the residuals depicting the difference between the observation data and the model.  For individual fit plots of the transit data, please see Figures \ref{fig:citiindividualplots} and \ref{fig:tessindividualplots}.}
\label{fig:AllDataGlobalPlot}
\end{figure}

The stacked TESS and citizen science light curves (Figure \ref{fig:AllDataGlobalPlot}) show excellent agreement with the best-fit model.  We can infer from the quality of the fit that the model is reliably fitting the citizen science and TESS data to derive the parameters of WASP-77 A b. The low uncertainty on transit depth of $0.013692 \pm2.3\times10^{-5}$ and the random distribution about 0 in the residuals implies a good fit between the data and the model. The ephemeris improvement with the addition of citizen science data further demonstrates the importance of utilizing all available data in conjunction with the continuous yield of TESS to fill in the gaps and keep exoplanet ephemerides fresh.  
%\citep{maxted2013,salz2015,turner2016,stassun2017,bonomo2017,cortes2020tramos,ivshinawinn2022,kokori2023}

\begin{deluxetable}{| p{2.0cm} | p{1.5cm} | p{1.9cm} | p{1.9cm} |p{1.7cm} | p{1.7cm}|}
\tabletypesize{\footnotesize}
\tablecaption{Parameters for comparison of global fits and joint transit + eclipse + RV analysis.  
\label{tab:globalfitcomparison}}
\tablehead{
\colhead{Parameter}& \colhead{Units} & \colhead{Citizen Science} & \colhead{TESS} & \colhead{Citizen Science+TESS} & \colhead{Transit+Eclipse+RV} 
}
\startdata
$P$ 
&[days] 
& 1.36002984 $\pm1.2\times10^{-7}$ 
& 1.36002877 $\pm1.9\times10^{-7}$ 
&  1.36002947 $\pm1.0\times10^{-7}$ 
& 1.360029395 $\pm5.7\times10^{-8}$ \\
\hline
$T_{mid}$ 
&[BJDTDB] 
& 2459957.33826 $\pm1.3\times10^{-4}$ 
& 2459169.880764 $\pm3.5\times10^{-5}$ 
& 2459957.337881 $\pm7.4\times10^{-5}$ 
& 2459957.337860 $\pm4.3\times10^{-5}$ \\
\hline
&[\(degree\)] 
& 90.00 $\pm1.1\times10^{-1}$ 
& 88.073 $\pm5.3\times10^{-2}$ 
& 88.371  $\pm5.9\times10^{-2}$ 
& 89.99 $\pm1.0\times10^{-1}$ \\
\hline
$R_p/R_*$  
&[\(ratio\)] 
& 0.11500 $\pm2.8\times10^{-4}$ 
& 0.11746 $\pm1.1\times10^{-4}$ 
& 0.117012 $\pm9.7\times10^{-5}$ 
& 0.117026 $\pm6.9\times10^{-5}$ \\
\hline
Transit light curves 
&\#
& 34
& 30
&64
& 64 transit, 6 eclipse\\
\hline
Radial velocity measurements 
& \#
&$n/a$ 
&$n/a$ 
&$n/a$
& 32\\
\hline
Propagated Mid-Transit to Jan 1 2030 
& [BJDTDB] 
& 2462503.31412 $\pm2.61\times10^{-4}$ ($\pm22.55 sec$) 
& 2462503.311279 $\pm4.67\times10^{-4}$ ($\pm40.35 sec$) 
& 2462503.313049 $\pm2.01\times10^{-4}$ ($\pm17.37 sec$) 
& 2462503.312887 $\pm1.26\times10^{-4}$ ($\pm9.94 sec$) \\
\enddata
\tablecomments{This table cross-references the ephemerides between the fits of only the terrestrial citizen science transit data, only TESS transit data, combined citizen science with Tess data, and the final joint fitting, which included both citizen science and TESS transit data with eclipse and radial velocity data.}
\end{deluxetable}

%%%%%%%%%%%%%%%%%%%%%%%%%%%%%%%%
%%         Conclusion         %%
%%%%%%%%%%%%%%%%%%%%%%%%%%%%%%%%
\section{Conclusion} \label{sec:conclusion}
A cornerstone of this effort is the joint-simultaneous analysis of transit photometry, eclipse photometry, and radial velocity measurements. By merging these data types, we are able to achieve a high level of precision in the best-fit model for WASP-77~A~b with a new orbital period improved to \(1.360029395 \pm 5.7 \times 10^{-8}\) days, mid-transit time to \(2459957.337860 \pm 4.3 \times 10^{-5}\) BJDTDB, mid-eclipse time to \(2459956.658192 \pm 6.7 \times 10^{-5}\) BJDTDB and mass to \(1.6654 \pm 4.5 \times 10^{-3} M_{J}\). Methodology from this study aids the community in updating exoplanet ephemerides precisely and showcasing the impact of collaborative data use and nested sampling \citep{pearson2022}.

The code for this study will be made publicly available on GitHub so that professional and citizen scientists can precisely update ephemerides of targets scheduled for observations by JWST and Ariel.  
With new innovations in obtaining spectroscopic measurements such as EXCITE (Exoplanet Climate Infrared Telescope) \citep{bernard2022design,tucker2018excite} or the Pandora SmallSat \citep{quintana2021pandora,hoffman2022pandora}, along with existing technologies, this code can be used by scientists for planning observations and performing atmospheric models that require the most precise and up to date ephemerides. 

The incorporation of citizen science data from sources such as Exoplanet Watch, Exoclock, and ETD in our study was an integral component of the achieved precision in the updating of WASP-77 A b's ephemeris giving us a broader time baseline. Furthermore, our results demonstrate that when citizen science light curve data is added to the TESS data it improves the propagated January 1, 2030 mid-transit time uncertainty by a factor of 2.3.  By incorporating citizen science observations into this study, we were able to leverage an observation taken on January 12, 2023 that would have not be possible without planned observations with space-based and large ground-based telescopes  or by using previously published mid-transit times alone.  With the number of newly discovered exoplanets continuing to grow there is an ongoing need for continued ephemeris maintenance.  \cite{dragomir20} demonstrates that exoplanets observed by TESS can have an over 30 minute uncertainty on the mid-transit time only one year after it was observed.  By supporting citizen science-based exoplanet initiatives, the astronomy community can keep up with the demand for exoplanet ephemeris updates.  

Our study underscores the significant impact of combining various data sources by including citizen science contributions, ground-based observations, space telescope data, and radial velocity measurements, for the refinement of exoplanetary parameters. This comprehensive approach not only enhances the precision of orbital parameters crucial for atmospheric characterization but also improves the observation scheduling for missions like JWST and Ariel.  The application of our publicly available Python code can be applied to various exoplanets, fitting radial velocity, transit, and eclipse data from both professional and citizen science sources substantially increasing ephemeris precision for the community. We advocate for sustained support of citizen science data contributions.  

\software{
    EXOTIC \citep{zellem2020}, 
    WOTAN \citep{HippkeWOTAN},
    Exotethys \citep{Morello2020},
    UltraNest \citep{ultranest2016}
}

\section{Acknowledgements}
Some of the research described in this publication was carried out in part at the Jet Propulsion Laboratory, California Institute of Technology, under a contract with the National Aeronautics and Space Administration. 

This research has used the NASA Exoplanet Archive and ExoFOP, which is operated by the California Institute of Technology, under contract with the National Aeronautics and Space Administration under the Exoplanet Exploration Program.

This publication uses data products from Exoplanet Watch, a citizen science project managed by NASA’s Jet Propulsion Laboratory on behalf of NASA’s Universe of Learning. 

This work is supported by NASA under award number NNX16AC65A to the Space Telescope Science Institute, in partnership with Caltech/IPAC, Center for Astrophysics—Harvard \& Smithsonian, and NASA Jet Propulsion Laboratory.

This research has used the NASA/IPAC Infrared Science Archive, which is funded by the National Aeronautics and Space Administration and operated by the California Institute of Technology.

This work is based in part on observations made with the Spitzer Space Telescope, which was operated by the Jet Propulsion Laboratory, California Institute of Technology, under a contract with NASA. NASA provided support for this work through an award issued by JPL/Caltech.

Based on observations collected at the European Southern Observatory under ESO programs 088.C-0011(A) and 1102.C-0744(C).

We thank the Ronald Greeley Planetary Scholarship at Arizona State University for supporting Suber Corley for this work.

The authors extend their gratitude to the Data and Analysis Center for Exoplanets (DACE) team at the University
of Geneva. As we embarked on this journey into exoplanet research utilizing radial velocity, their DACE Python
package, online graphical tool, and accompanying tutorials proved invaluable in helping us grasp the nuances of radial
velocity data analysis. We particularly appreciate their prompt and unwavering support when we reached out for
assistance. Their willingness to share expertise and enrich the broader community is exemplary and we are truly
thankful for that.

This research has made use of NASA's Astrophysics Data System. Some/all of the data presented in this paper were obtained from the Mikulski Archive for Space Telescopes (MAST). STScI is operated by the Association of Universities for Research in Astronomy, Inc., under NASA contract NAS5-26555. Support for MAST for non-HST data is provided by the NASA Office of Space Science via grant NNX13AC07G and by other grants and contracts. 

Some/all of the data presented in this paper were obtained from the Mikulski Archive for Space Telescopes (MAST) at the Space Telescope Science Institute. The specific observations analyzed can be accessed via \dataset[doi:10.17909/t9-nmc8-f686]{https://archive.stsci.edu/doi/resolve/resolve.html?doi=10.17909/t9-nmc8-f686}, 
\dataset[doi:10.17909/T97P46]{https://archive.stsci.edu/doi/resolve/resolve.html?doi=10.17909/T97P46},
\dataset[doi:10.17909/3fmp-zj55]{https://archive.stsci.edu/doi/resolve/resolve.html?doi=10.17909/3fmp-zj55}.

STScI is operated by the Association of Universities for Research in Astronomy, Inc., under NASA contract NAS5–26555. Support to MAST for these data is provided by the NASA Office of Space Science via grant NAG5–7584 and by other grants and contracts.

This research has made use of the SIMBAD database, operated at CDS, Strasbourg, France. 

We acknowledge with thanks the transit observations from the AAVSO International Database contributed by observers worldwide and used in this research. 

This research has made use of the VizieR catalogue access tool, CDS, Strasbourg, France. 

Observation taken by Jake Postiglione was funded by PSC-CUNY Award 65172-00 53

Based on observations collected at the European Organisation for Astronomical Research in the Southern Hemisphere under ESO programme(s) 088.C-0011(A), and 0102.C-0618(A). 

Based on data obtained from the ESO Science Archive Facility with DOI(s): \url{https://doi.org/10.18727/archive/33}

MicroObservatory is maintained and operated as an educational service by the Center for Astrophysics Harvard and Smithsonian and is a project of NASA’s Universe of Learning, supported by NASA Award NNX16AC65A. Additional MicroObservatory sponsors include the National Science Foundation, NASA, the Arthur Vining Davis Foundations, Harvard University, and the Smithsonian Institution.

We thank Frank Sienkiewicz of the Science Education department, CfA, who maintains and schedules the MicroObservatory telescopes for exoplanet observations.

P. S. acknowledges support provided by NASA through the NASA FINESST grant 80NSSC22K1598.

\bibliography{sample631}{}
\bibliographystyle{aasjournal}

\pagebreak

\end{document}